\documentclass[review,5p,times]{elsarticle}
\pdfoutput=1
\usepackage{etex}
\usepackage{color}
\usepackage{colortbl}
\usepackage{lscape}
\usepackage{amsmath}
\usepackage{txfonts}
\usepackage{marvosym}
\usepackage[lined,boxed,linesnumbered]{algorithm2e}
\usepackage[ansinew]{inputenc}
\usepackage[french, english]{babel}
\usepackage[T1]{fontenc}

\usepackage{lmodern}  \normalfont
\usepackage{hyperref}
\usepackage{algorithmic}
\algsetup{linenosize=\footnotesize\ttfamily}

\usepackage{amsfonts}
\usepackage{amssymb}
\usepackage{ae,aecompl}
\usepackage{epsfig}

\usepackage{booktabs}
\usepackage{graphics}
\usepackage{array}
\usepackage{graphicx}
\usepackage{tikz}

\usepackage{tabularx}
\usepackage{listings}
\usepackage{longtable}
\usepackage{lineno}

\usepackage{mathrsfs}
\usepackage{lettrine}
\usepackage{tabularx}
\usepackage{epsfig, floatflt, amssymb} 
\usepackage{moreverb} 
\usepackage{cases}  
\usepackage{multirow} 
\usepackage{url} 
\usepackage[all]{xy} 
\usepackage{textcomp} 
\usepackage[right]{eurosym}
\usepackage{setspace} 

\usepackage{epic,eepic}
\usepackage{soul}
\usepackage[nottoc]{tocbibind} 
\usepackage{appendix}
\usepackage{subfig}
\usepackage{ifthen}
\usepackage{xkeyval}
\usepackage{xfor}
\usepackage{amsgen}
\usepackage{etoolbox}
\usepackage{longtable} 
\usepackage{supertabular}
\usepackage{datatool}

\usepackage{fancybox}
\usepackage[leftcaption]{sidecap}
\usepackage[labelsep=endash, textfont={footnotesize, singlespacing}, margin=10pt, format=plain, labelfont=bf]{caption}
\usepackage[Conny]{fncychap} 

\definecolor{name}{rgb}{0.5,0.5,0.5}
\definecolor{javared}{rgb}{0.6,0,0} 
\definecolor{javagreen}{rgb}{0.25,0.5,0.35} 
\definecolor{javapurple}{rgb}{0.5,0,0.35} 
\definecolor{javadocblue}{rgb}{0.25,0.35,0.75} 

\newcommand{\fakeparagraph}[1]{\smallskip\noindent\textbf{#1.}}

\newcommand*\circled[1]{\tikz[baseline=(char.base)]{
            \node[shape=circle,draw,inner sep=0.5pt, minimum size=0.5pt] (char) {#1};}}

\rmfamily
\DeclareFontShape{T1}{lmr}{bx}{sc}{<->ssub * cmr/bx/sc}{}
\DeclareSymbolFont{calletters}{OMS}{cmsy}{b}{n}
\DeclareSymbolFontAlphabet{\mathcal}{calletters}
\DeclareSymbolFont{rmletters}{OMS}{ptm}{m}{n}
\DeclareSymbolFontAlphabet{\mathrm}{rmletters}

\usepackage{mathrsfs}
\usepackage{lettrine}
\usepackage{tabularx}
\usepackage{epsfig, floatflt, amssymb} 
\usepackage{moreverb} 
\usepackage{cases}  
\usepackage{multirow} 
\usepackage{url} 
\usepackage[all]{xy} 
\usepackage{textcomp} 
\usepackage[right]{eurosym}
\usepackage{setspace} 

\usepackage{epic,eepic}
\usepackage{soul}
\usepackage[nottoc]{tocbibind} 
\usepackage{appendix}
\usepackage{subfig}
\usepackage{ifthen}
\usepackage{xkeyval}
\usepackage{xfor}
\usepackage{amsgen}
\usepackage{etoolbox}
\usepackage{longtable} 
\usepackage{supertabular}
\usepackage{datatool}

\usepackage{fancybox}
\usepackage[leftcaption]{sidecap}

\usepackage[labelsep=endash, textfont={footnotesize, singlespacing}, margin=10pt, format=plain, labelfont=bf]{caption}
\usepackage[Conny]{fncychap} 

\makeatletter

\renewcommand{\SC@figure@vpos}{c}
\renewcommand{\fnum@figure}{\small\textbf{Figure~\thefigure}}
\renewcommand{\fnum@table}{\small\textbf{Table~\thetable}}

\begin{document}

\begin{frontmatter}
\author{Saurabh Chauhan}
\ead{saurabh.bharatsinh-chauhan@in.abb.com}
\author{Pankesh Patel}
\ead{pankesh.patel@in.abb.com}

\address{ABB Corporate Research, Bangalore, India}

\title{An IoT application development using IoTSuite}

\begin{abstract}
Application development in the Internet of Things~(IoT) is challenging because it involves dealing 
with issues that attribute to different life-cycle phases. 
First, the application logic has to be analyzed and then separated into a set of distributed tasks for 
an underlying network. Then, the tasks have to be implemented for the specific hardware. Moreover, we take different IoT applications and present development of these applications using IoTSuite.

In this paper, we introduce a design and implementation of ToolSuite, a suite of tools, for reducing burden of each 
stage of IoT application development process. We take different
class of IoT applications, largely found in the IoT literature, and 
demonstrate these IoT application development using IoTSuite. 
These applications have been tested on several IoT technologies
such as Android, Raspberry PI, Arduino, and JavaSE-enabled devices, Messaging protocols such as MQTT, CoAP, WebSocket, Server technologies such as Node.js, Relational database such as MySQL, and Microsoft Azure Cloud services.

\begin{keyword}
Internet of Things, Wireless Sensor Networks, Ubiquitous/Pervasive Computing, Programming Framework, Toolkit, Domain-specific languages, Development Life-cycle
\end{keyword}
\end{abstract}
\end{frontmatter}

\section{Introduction}

Recent technological advances in computer and communication technology 
have been fueling a tremendous growth in a number of smart
objects (or \emph{things})~\cite{vasseur2010interconnecting}.
In the Internet of Things\footnote{\url{http://www.grifs-project.eu/data/File/CASAGRAS\%20FinalReport\%20(2).pdf}}, 
these ``things'' acquire intelligence, thanks to the fact that they access information 
that has been aggregated by other things.  For example, a building interacts 
with its residents and surrounding buildings in case of fire for safety and security of residents, 
offices adjust themselves automatically accordingly to user preferences while minimizing energy consumption, 
or traffic signals control in-flow of vehicles according to the current highway status~\cite{de2009internet,IoTsurvey}. 
It is the goal of our work to enable the development of such applications. 
In the following, we discuss one of such applications. 

\subsection{Challenges and contributions}
Application development in the IoT is challenging because stakeholders\footnote{we will use the term stakeholders as
used in software engineering to mean people, who are involved in the application development.
Examples of stakeholders defined in~\cite{softwareArchtaylor2010}  are software designer, developer, domain
expert, technologist, etc.} have to address issues that are attributed 
to different life cycles phases, including development, deployment, and maintenance~\cite{bischoff2007life}. 
At the \textbf{\emph{development phase}}, the application logic has to be analyzed and separated into a set of distributed 
tasks for the underlying network consisting of a large number of heterogeneous entities. 
Then, the tasks have to be  implemented for  the specific platform of a device. At the \textbf{\emph{deployment phase}}, the application logic has to be deployed onto a large number of devices. Manual effort in above two phases for hundreds to thousands of heterogeneous devices is a time-consuming and error-prone process.   

An important challenge that needs to be addressed in the IoT is to enable the rapid development 
of applications with minimal effort by stakeholders involved in the process~\cite{patel:hal-00788366, DBLP:journals/corr/PatelLB16}.
In our previous publications~\cite{patelhalscube,  Chauhan:2016:DFP:2897035.2897039, patel-icse14}, we have provided a complete tour of our
IoT application development process, summarized in Section~\ref{sec:ourapproach}. 
This paper goes beyond it. In particular, it describes implementation technologies, tools, language used and their 
rationales of choosing them~\cite{patel-comnet-iot15, 7460669, 7467275}. Although various efforts exist in literature for making IoT 
application development easier, very few of them are publicly available for stakeholders to 
choose from. Given the usefulness of open source, our aim is to provide an opportunity 
to community for the creation of novel software engineering tools and the conduction 
of novel research for IoT application development.  

The main contribution of this paper is an implementation of ToolSuite, a suite of tools,
for reducing burden at different phases for IoT application 
development~(detail in Section~\ref{sec:components}). Moreover, we take 
different class of IoT applications~\cite{appdevIoTpatel2011} and describe
an application development process using IoTSuite. 

\textbf{Outline.} The remainder of this paper is organized as follows: Section \ref{sec:ourapproach} presents the development framework that includes the proposed modeling languages and automation techniques. Section \ref{sec:components} presents the implementation of the development framework and describes implementation technologies, tools, and languages used. Section \ref{sec:appdev} describes step by step IoT application development process using IoTSuite. It also focus on different class of application to demonstrate application development using IoTSuite. Finally, in Section
~\ref{sec:conclusion}, we conclude this article and briefly mention some of future directions. 
\section{IoT application development process}\label{sec:ourapproach}

This section presents our development framework that separates IoT application development into different concerns, namely \emph{domain}, \emph{platform}, \emph{functional}, and \emph{deployment}. It integrates a set of high-level modeling languages to specify such concerns.  These languages are supported by automation techniques at various phases of application development process. Stakeholders carry out the following steps in order to develop an IoT using our approach:

\begin{figure*}[!ht]
\centering
\includegraphics[width=0.7\textwidth]{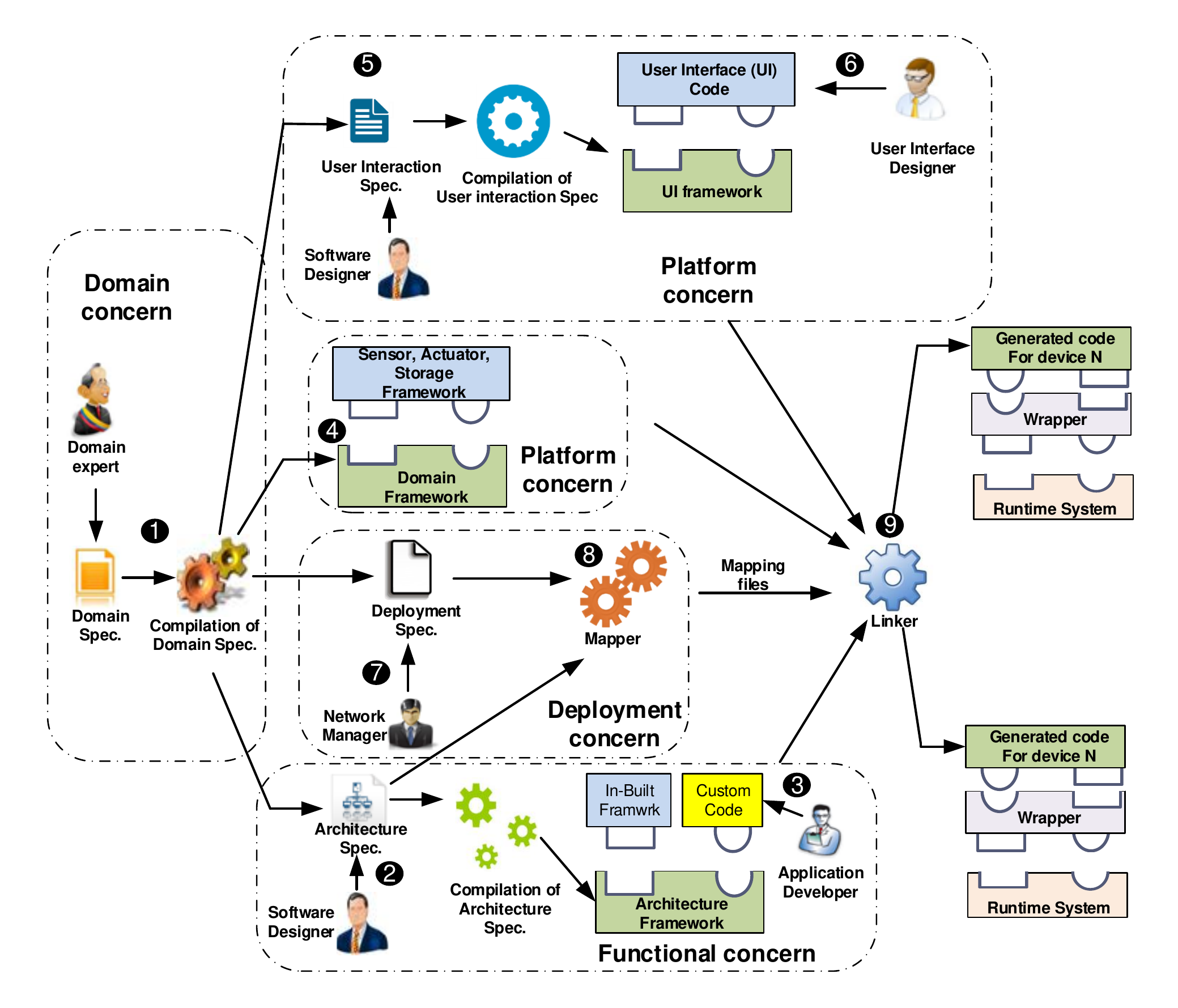}
\caption{IoT application development: the overall process}
\label{fig:devcycle}
\end{figure*}

\subsection{Domain Concern} 

This concern is related to concepts that are specific to a domain (e.g., building automation, transport) of an IoT. The stakeholders task regarding such 
concern consists of the following step:

\fakeparagraph{\em Specifying and compiling domain specification} 
The domain expert specifies a domain specification using the Domain Language~(DL) 
(Step~\circled{1} in Figure~\ref{fig:devcycle}). The domain specification includes specification of resources, which are responsible for interacting with Entities of Interest~(EoI). This includes \emph{tags}~(identify EoI), \emph{sensors}~(sense EoI), \emph{actuators}~(affect EoI), and \emph{storage}~(store information about EoI). In the domain specification, resources  are specified in a high-level manner to abstract low-level details from the domain expert.
\subsection{Functional Concern}
This concern is related to concepts that are specific to functionality of an IoT application. An example of a functionality is to open a window when an average temperature value of a room is greater than $30^{\circ} C$. The stakeholders task regarding such 
concern consists of the following steps: 

\fakeparagraph{\em Specifying application architecture} 
Referring the domain specification, the software designer specifies an application architecture using the Architecture Language~(AL)-(Step \circled{2} in Figure~\ref{fig:devcycle}). It consists of specification of computational 
services and interaction among them. A computational service is fueled by sensors and storage defined in the domain specification. They process inputs data and take appropriate decisions by triggering actuators defined in the domain specification. The application architecture  consists of \textit{common} operations and \textit{custom} components that are specific to the application logic.

\fakeparagraph{\em Implementing application logic} 
The compilation of an architecture specification generates an architecture framework~(Step~\circled{3} 
in Figure~\ref{fig:devcycle}).  The architecture framework contains abstract classes, 
corresponding to  each  computational service, that hide interaction details with other 
software components  and allow the application developer to focus only  on application logic. 
The application developer implements only abstract methods of generated abstract classes, described in our work~\cite[p.~73]{Patel201562}.
We have integrated a framework for common operations. This further reduces the development effort  for commonly found operations in IoT application and provides re-usability. 

\subsection{Platform Concern}
This concern specifies the concepts that fall into computer 
programs that act as a translator between a hardware device and an application. 
The stakeholders task regarding such concern consists of the following steps:
		
\fakeparagraph{\em Generating device drivers}
The compilation of domain specification generates a domain framework~(Step~\circled{4} in Figure~\ref{fig:devcycle}). 
It contains {\em concrete classes} corresponding to concepts defined in the domain specification. The concrete classes contain concrete
methods to interact with other software components and platform-specific device drivers,
described in our work~\cite[p.~75]{Patel201562}. We have integrated existing open-source sensing framework\footnote{\url{http://www.funf.org/}} 
for Android devices. Moreover, we have implemented sensing and actuating framework
for Raspberry Pi and storage framework for MongoDB, MySQL, and Microsoft AzureDB. So, the device developers do not have to implement platform-specific sensor, actuator, and storage code.

\fakeparagraph{\em Specifying user interactions}
To define user interactions, we present a set of {\em abstract interactors}, similar to work~\cite{Balland-2013}, that denotes information exchange between an application and a user. The software designer specifies them using User Interaction Language~(UIL)~(Step~\circled{5} in Figure~\ref{fig:devcycle}).

\fakeparagraph{\em Implementing user-interface code}
Leveraging the user interaction specification, the development framework generates a User Interface~(UI) framework~(step~\circled{6} in Figure~\ref{fig:devcycle}). The UI framework contains a set of {\em interfaces} and {\em concrete classes} corresponding to resources defined in the user interaction specification.  The concrete classes contain concrete methods for interacting with other software components. 
The user interface designer implements {\em interfaces}. These interfaces implement code that connects appropriate UI elements to concrete methods. 
For instance, a user initiates a command to heater by pressing UI element such as \texttt{button} that invokes a \texttt{sendCommandToHeater()} concrete method.
 
\subsection{Deployment Concern}

This concern is related to deployment-specific concepts that describe the
information about a device and its properties placed in the target deployment. 
It consists of the following steps:

\fakeparagraph{\em Specifying target deployment} 
Referring the domain specification, the network manager describes a deployment specification using the Deployment Language~(DL)~(Step~\circled{7} in Figure~\ref{fig:devcycle}). The deployment specification 
includes the details of each device as well as abstract interactors specified in the
user interaction specification.

\fakeparagraph{Mapping} The mapper takes a set of devices defined in the deployment specification and  a set of computation components defined in the architecture specification(Step-~\circled{8} in Figure~\ref{fig:devcycle}). It maps computational service to a device.  The current version of mapper algorithm~\cite{Patel201562} selects devices randomly and allocates computational services to the selected devices.

\subsection{Linking}
The linker combines the code generated by various stag-es and creates packages that can be deployed on devices~(St-ep~\circled{9} in Figure~\ref{fig:devcycle}).  This stage supports the application deployment phase by producing device-specific code to result in a distributed software system collaboratively hosted by individual devices, thus providing automation at the deployment phase.

The final output of linker is composed of three parts: (1) a \emph{runtime-system} runs on each individual device and provides a support for executing distributed tasks, (2) a \emph{device-specific code} generated by the linker module, and (3) a \emph{wrapper} separates generated code from the linker module and underlying runtime system by implementing interfaces. 
\section{Components of IoTSuite}\label{sec:components}

\begin{figure*}[!ht]
	\centering \includegraphics[width=0.8\linewidth]{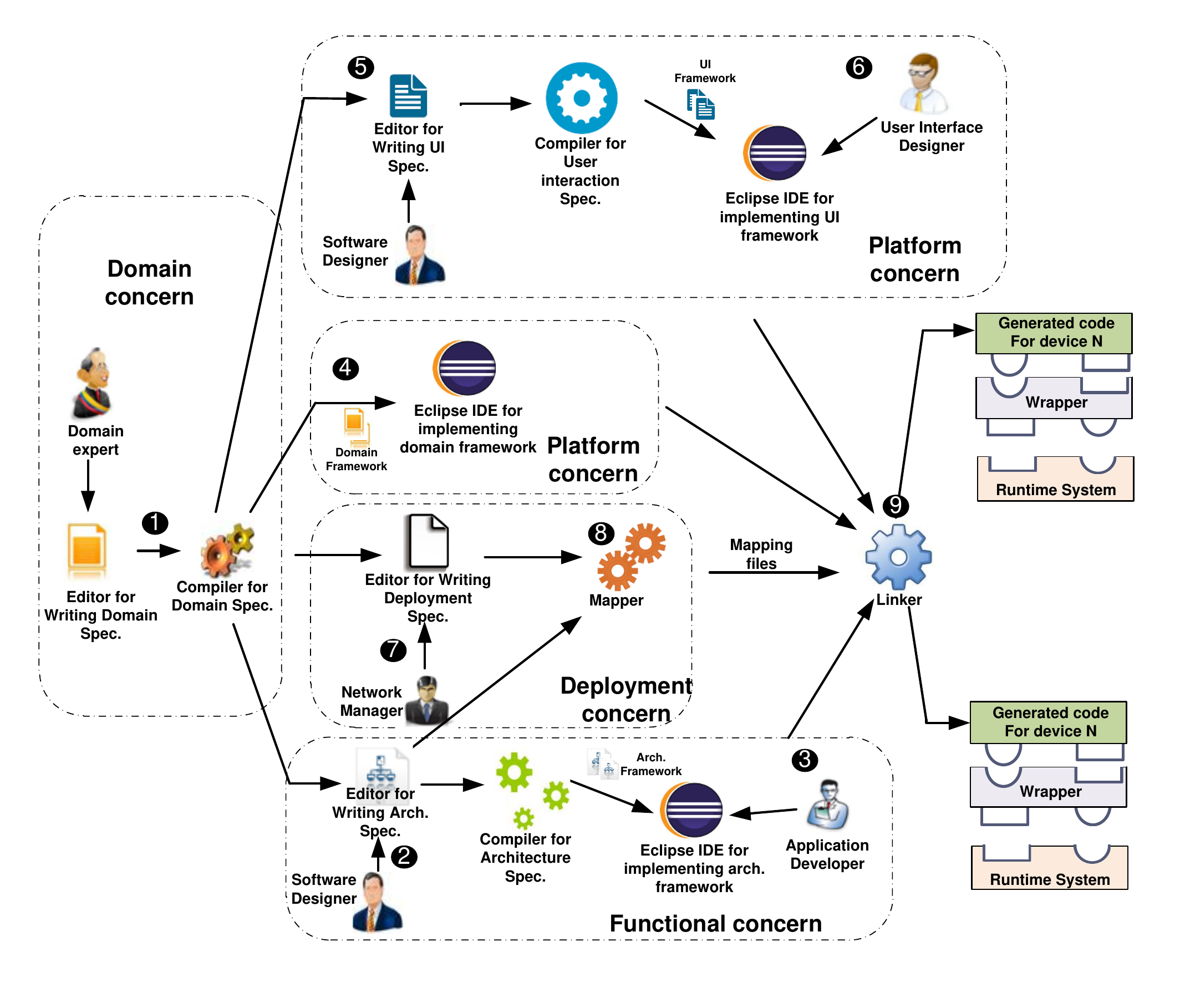} 
	\caption{Overview of components in IoTSuite.} \label{fig:srijansuite}
\end{figure*}

This section presents the implementation of  the proposed IoT application development process discussed in Section~\ref{sec:ourapproach}.  
In particular, it describes implementation technologies, tools, language used  
and their rationales of choosing them. Figure~\ref{fig:srijansuite} shows the various components 
at each phase of application development that stakeholders can use, described below.
 
\begin{itemize}
	\item \textbf{Editor:} It helps stakeholders to write high-level specifications, including domain, architecture, userinteraction, and deployment specification.  
	\item \textbf{Compiler:}	It parses the high-level specifications and translates them 
		   into the code that can be used by other components in the system. 
	\item \textbf{Mapper:} It maps computational services described in an architecture specification 
		  to devices listed in an deployment specification.	
	\item \textbf{Linker:} It combines and packs code generated by various stages of 
	compilation into packages that can be deployed on devices. 
	\item  \textbf{Runtime system: }It is responsible for a distributed execution of an application. 
\end{itemize}

Each component is described in detail in the following sections.

\subsection{Editor}\label{sec:editor}
The editor provides supports for specifying high-level textual 
languages with the facilities of outline view, syntax coloring, code folding, error checking, rename re-factoring, and auto completion. 
The editor support is provided at different phases of IoT application development 
to help stakeholders  illustrated in Figure~\ref{fig:srijansuite}: 
(1) editor for specifying a domain to aid the domain expert, (2) editor 
for specifying an application architecture to aid the software designer, (3) editor for specifying an userinteraction to aid the software designer and, (4) editor for specifying a deployment scenario to aid the network manager. 
\begin{figure*}[!ht]
	\centering \includegraphics[width=1.0\linewidth]{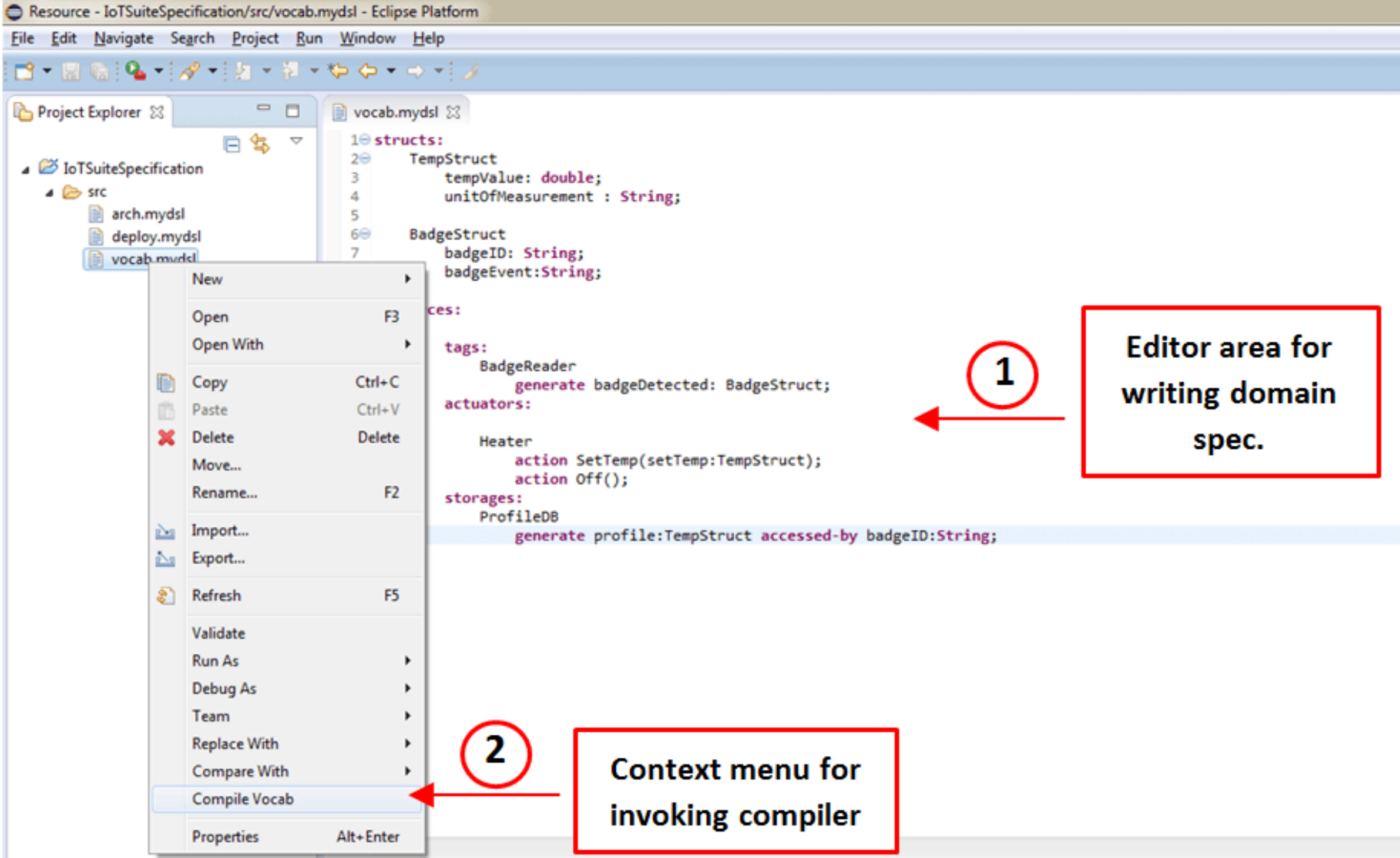} 
	\caption{Editor support for writing domain specification in IoTSuite.} \label{fig:editor}
\end{figure*}

	\begin{figure*}[!ht]
\centering
\includegraphics[width=1.0\linewidth]{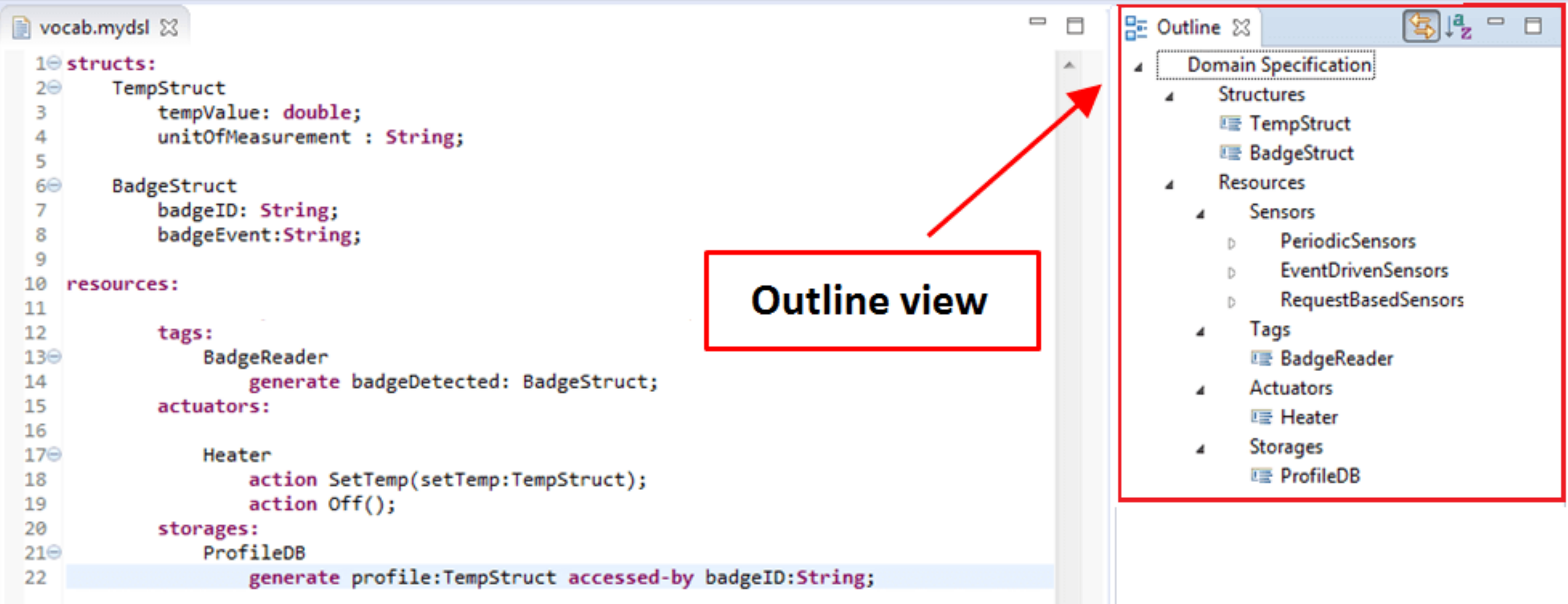}
\caption{IoTSuite editor feature: Outline view}
\label{fig:outline}
\end{figure*}

We take the editor for domain specification as an example to demonstrate 
an editor support provided by IoTSuite, illustrated in Figure~\ref{fig:editor}. 
The zone~\circled{1} shows the editor, where the domain expert writes a domain specification. 
The zone~\circled{2} shows the context menu, where the domain expert invokes the  
compiler for domain specification to generate a  framework.

\fakeparagraph{Features of editor}  We use  Xtext\footnote{\url{http://www.eclipse.org/Xtext/}} 
for a full fledged editor support, similar to work in~\cite{bertran2012diasuite}. 
The Xtext  is a framework for a development of domain-specific languages, 
and provides an editor with features such as \textit{syntax coloring}, \textit{error checking}, \textit{auto completion}, \textit{rename re-factoring}, \textit{outline view}, and \textit{code folding}:
\begin{figure*}[!ht]
\centering
\includegraphics[width=1.0\textwidth]{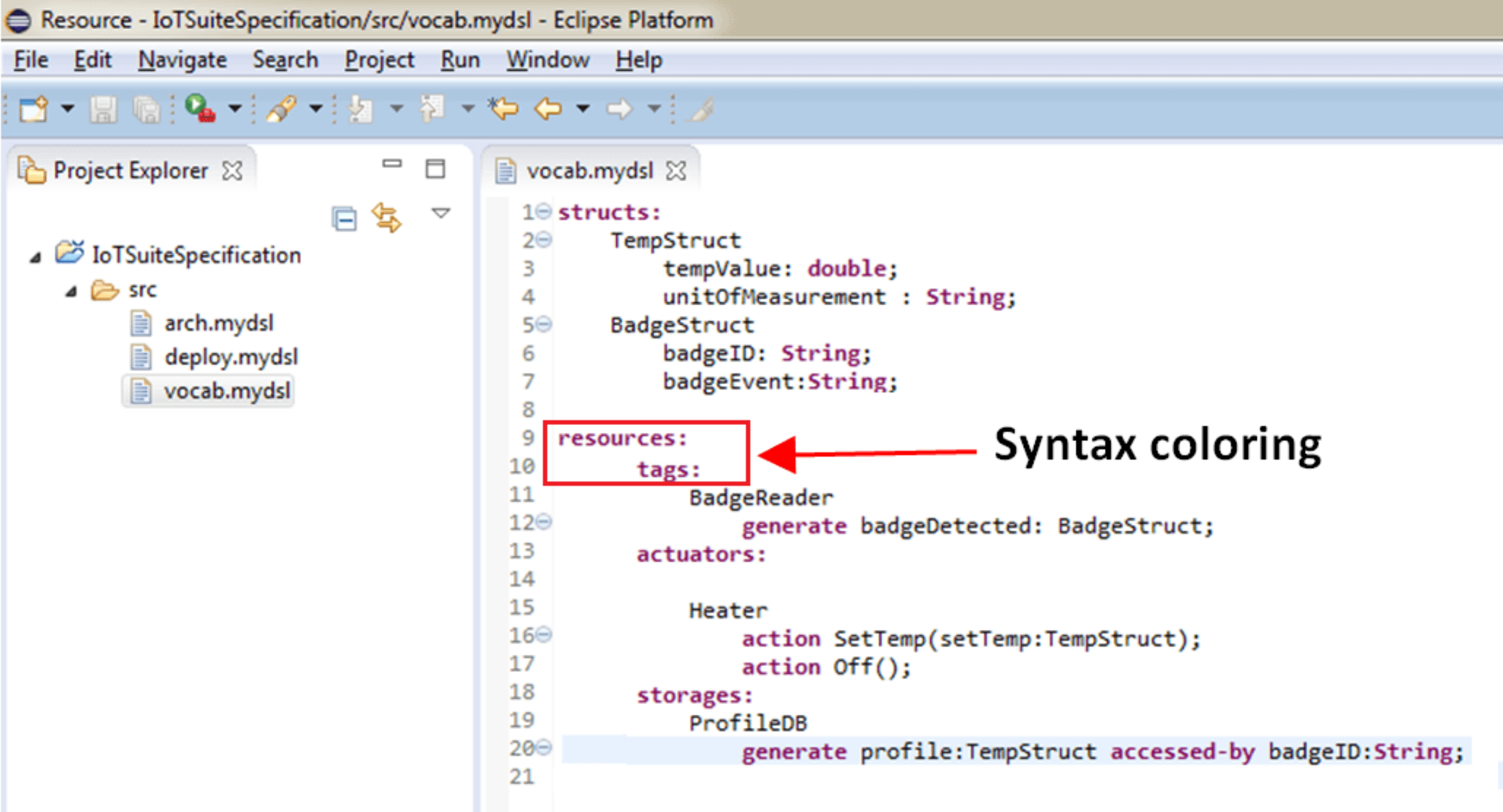}
\caption{IoTSuite editor feature: Syntax coloring}
\label{fig:coloring}
\end{figure*}

\begin{itemize}
	\item We implemented \textit{Outline/Structure view} feature, wh-ich is displayed on top most right side of the screen~-(Refer~Figure~\ref{fig:outline}). It displays an outline of a file highlighting its structure. This is useful for quick navigation. As shown in Figure~\ref{fig:outline}, \texttt{vocab.mydsl} file contains a large number of  structures, sensors,  and actuators, than from outline view by just clicking on  particular structure (e.g., \texttt{TempStruct}) it  navigates to \texttt{TempStruct} definition in the \texttt{vocab.mydsl}. So, developers don't need to look into an entire file.

\item Using \textit{syntax coloring} feature, keywords, comments, and other datatype elements are appeared in colored text. Here, \texttt{resources}, and \texttt{tags} are appeared in colored text as shown in~Figure~\ref{fig:coloring}.

\begin{figure*}[!ht]
\centering
\includegraphics[width=1.0\textwidth]{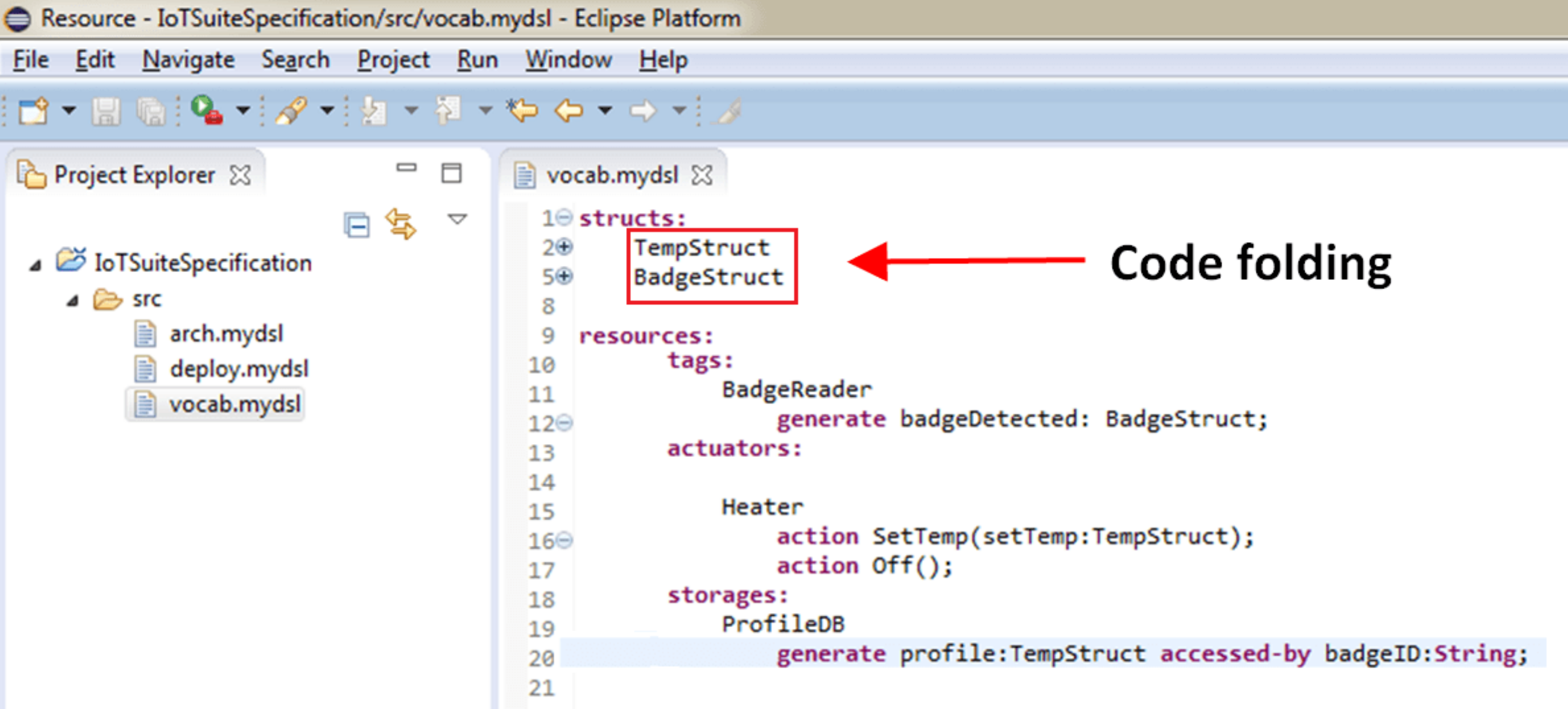}
\caption{IoTSuite editor feature: Code folding}
\label{fig:folding}
\end{figure*}

\item Using \textit{code folding}, developer can collapse parts of a file that are not important for current task. In order to implement code folding, click on dashed sign located in left most side of the editor. When developer clicked on dashed sign, it will fold code and sign is converted to plus. In order to unfold code, again click on plus sign.
As shown in Figure~\ref{fig:folding}, the code is folded for \texttt{TempStruct}, and \texttt{BadgeStruct}. 

\item The \textit{error checking} feature guides developer if any error is there.  An error in the file is marked automatically e.g., violation of the specified syntax or reference to undefined elements. Error checking indicates if anything is missing/wrong  in particular specification file.

\begin{figure*}[!ht]
\centering
\includegraphics[width=1.0\textwidth]{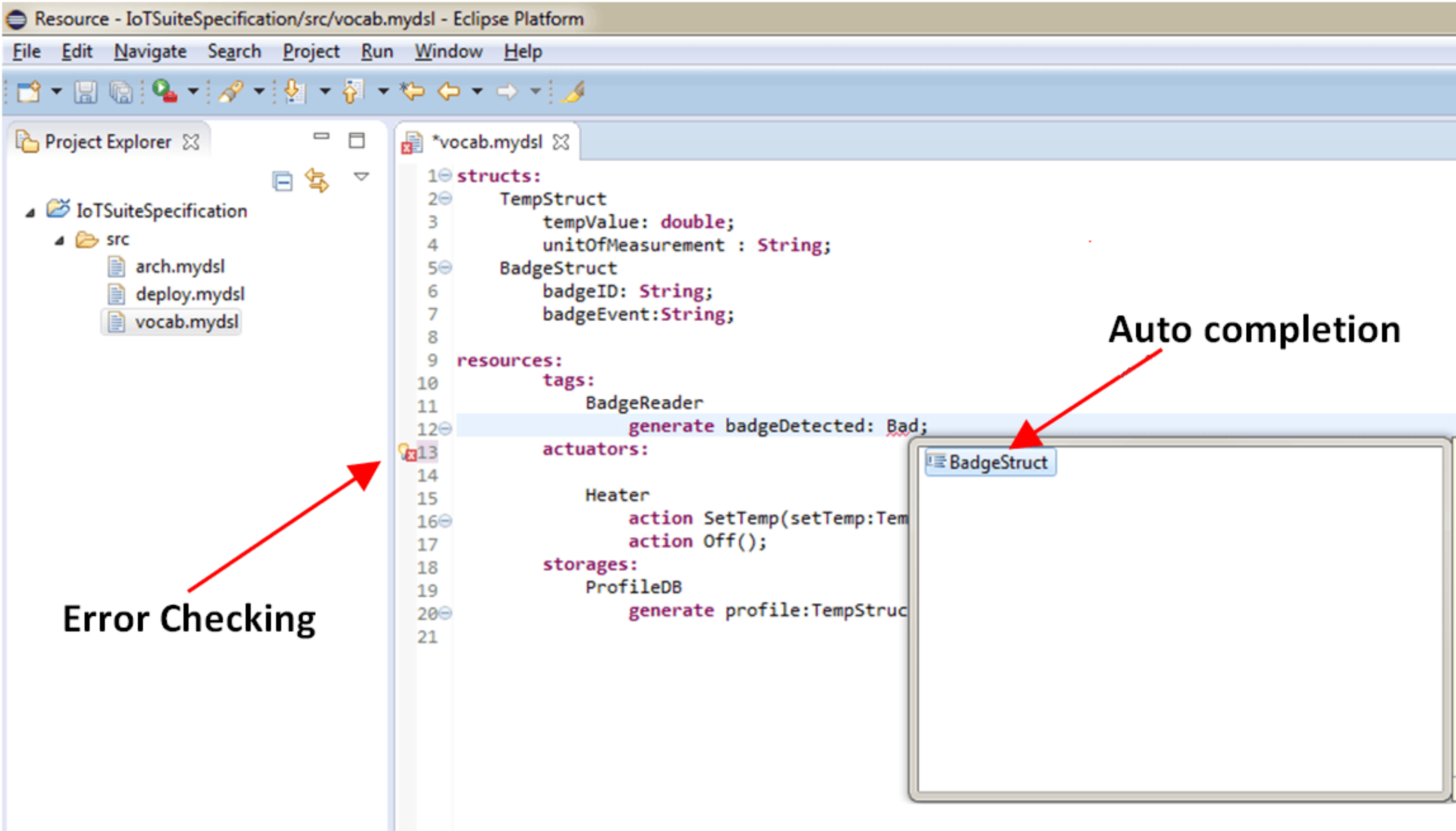}
\caption{IoTSuite editor features: Error checking \& Auto completion}
\label{fig:error}
\end{figure*}

\item The \textit{Auto completion} is used to speed up writing text in specification files. In order to use auto completion feature, developer needs to press ctrl+space key at current cursor position,  so it will provide suggestion. Here in \texttt{BadgeReader} definition, if developer writes Bad and press ctrl+space than it will suggest developer to write \texttt{BadgeStruct}~(Refer~Figure~\ref{fig:error}).
\end{itemize}

\subsection{Compiler}\label{sec:compiler}
The compiler parses high-level specifications and translates them into code that
can be used by other components in the system. This component is composed of two modules: 
(1) \emph{parser}. It converts high-level specifications into data structures that can be 
used by the code generator. (2) \emph{code generator}. It uses outputs of the parser and 
produces files in a target implementation language. In the following, each of these modules are discussed.
  \begin{figure*}[!ht]
	\centering \includegraphics[width=1.0\linewidth]{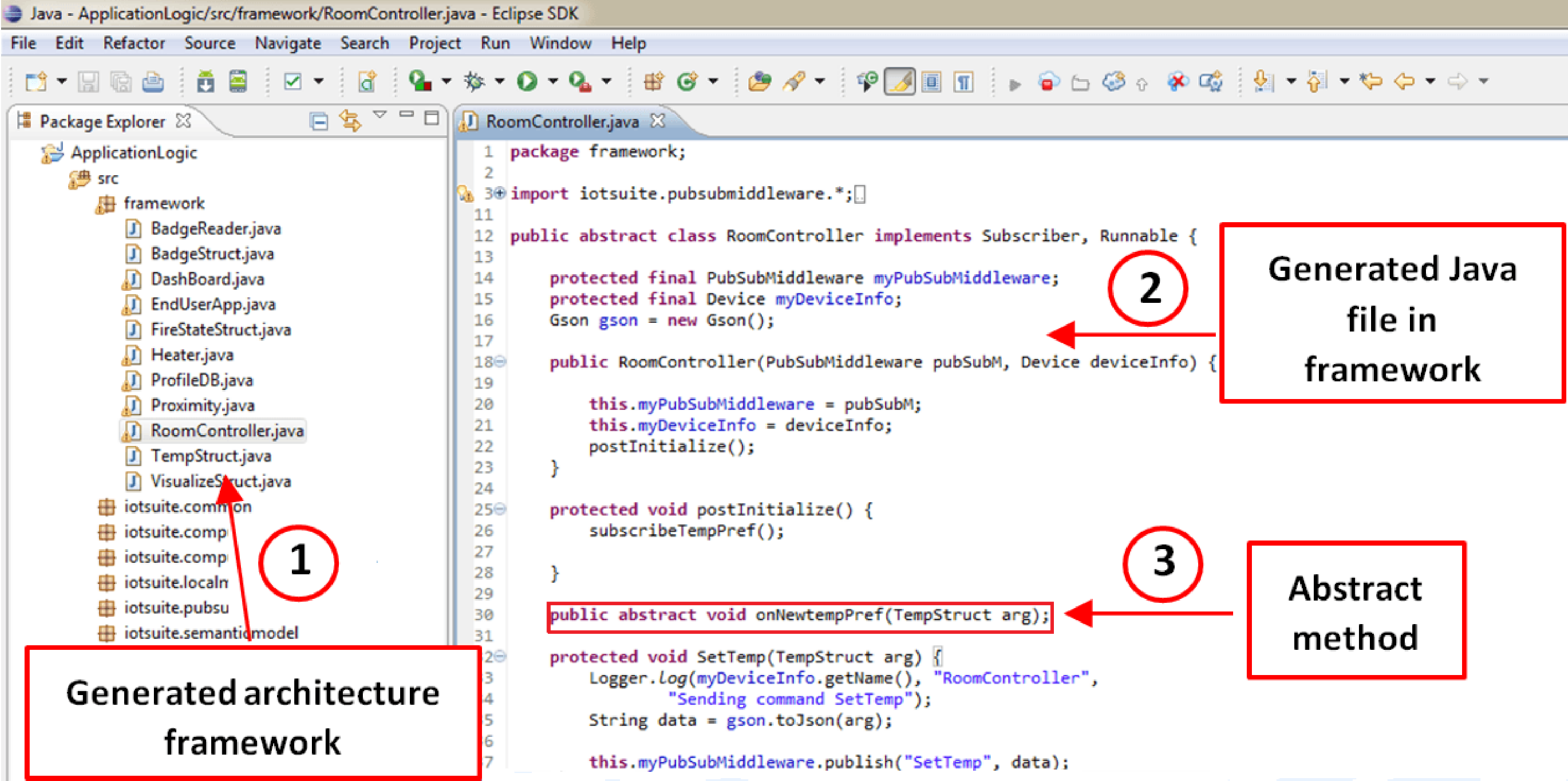} 
	\caption{Generated architecture framework in Eclipse} \label{fig:eclipsegeneratedframework}
\end{figure*} 

\fakeparagraph{Parser} It converts high-level specifications~(domain, architecture, userinteraction, and deployment specification) 
into data structures that can be used by the code generator. 
Apart from this core functionality, it also checks syntax of specifications and reports errors to stakeholders. 
The parser is implemented using ANTLR parser generator~\cite{parr2007definitive}.
The ANTLR parser is a well-known parser generator that creates parser files from grammar descriptions.


\fakeparagraph{Code generator}
Based on parser outputs, the code generator creates required files. It is composed of two sub-modules: 
(1) \emph{core-module}, (2) \emph{plug-in}.  The core-module manages a repository of plug-ins. 
Each plug-in is specific to a target implementation code. 
The target code could be in any programming  language~(e.g., Java, Python).  
Each plug-in is defined as template files, which the core-module uses to generate code. 
The key advantage of separating core-module and plug-in is that it simplifies 
an implementation of a new code generator for a target implementation.

The plug-ins are implemented using StringTemplate Engine\footnote{\url{http://www.stringtemplate.org/}}, a Java 
template engine for generating source code or any other formatted text output. 
In our prototype implementation, the target code is in the Java programming language compatible with Eclipse IDE. 
However, the code generator is flexible  to generate code in any 
object-oriented programming language, thanks to the architecture of the  
code generator that separates core-module and plug-ins. 

We build two compilers to aid stakeholders shown in Figure~\ref{fig:srijansuite}. 
(1) compiler for a domain specification. It translates a domain specification 
and generates a domain framework, and a customized architecture and deployment grammar to aid stakeholders. 
(2) compiler for an architecture specification. It translates an architecture specification 
and generates an architecture framework to aid the application developer. 
The both generated frameworks are compatible with Eclipse IDE. For example, 
Figure~\ref{fig:eclipsegeneratedframework} shows a generated architecture framework 
containing Java files~(\circled{1} in Figure~\ref{fig:eclipsegeneratedframework}) in Eclipse IDE. 
\circled{2} in Figure~\ref{fig:eclipsegeneratedframework} shows a generated Java file in the architecture 
framework for a \texttt{RoomController}.  Note that the generated framework contains abstract 
method~(\circled{3} in Figure~\ref{fig:eclipsegeneratedframework}), which are 
implemented by the  application developer using Eclipse IDE.
\subsection{Mapper}\label{sec:mapper}
The mapper produces a mapping from a set of computational services to a set of devices.
\begin{figure*}[!ht]
	\centering \includegraphics[width=1.0\linewidth]{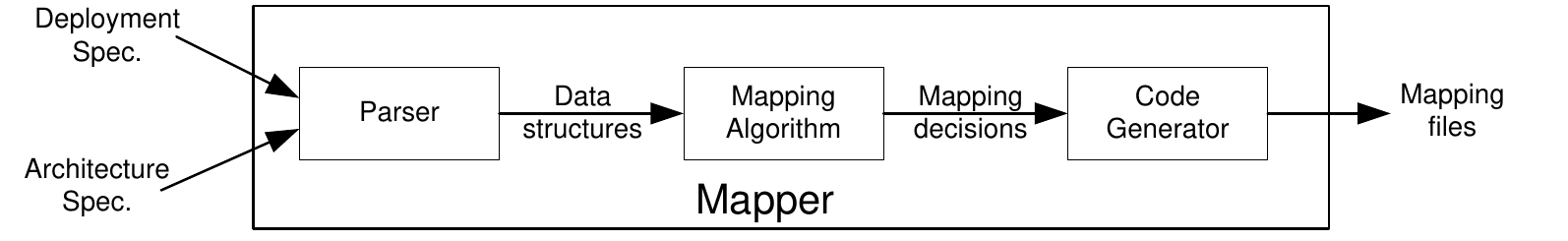} 
	\caption{Architecture of the mapper component in IoTSuite.}\label{fig:mappercomponent}
\end{figure*}
Figure~\ref{fig:mappercomponent}  illustrates the architecture of the mapper component. 
This component parses a deployment and architecture specification. 
The parser converts high-level specifications into appropriate 
data structures that can be used by the mapping algorithm. The mapping algorithm maps  
computational services described in the architecture specification to devices described 
in the deployment specification and produces mapping decisions into appropriate data structures. 
The code generator consumes the data structures and generates mapping files that 
can be used by the linker component.

In our current implementation, this module randomly maps computational services to a set of devices. 
However, due to generality of our framework, more sophisticated mapping 
algorithm can be plugged into the mapper component.

\subsection{Linker}\label{sec:linker}
The linker combines and packs code generated by various stages of 
compilation into packages that can be deployed on devices. It merges a 
generated architecture framework, application logic, mapping code, device drivers, 
and domain framework. This component supports the deployment phase by 
producing device-specific code to result in a distributed software system 
collaboratively hosted by individual devices.

\begin{figure*}[!ht]
	\centering \includegraphics[width=0.85\linewidth]{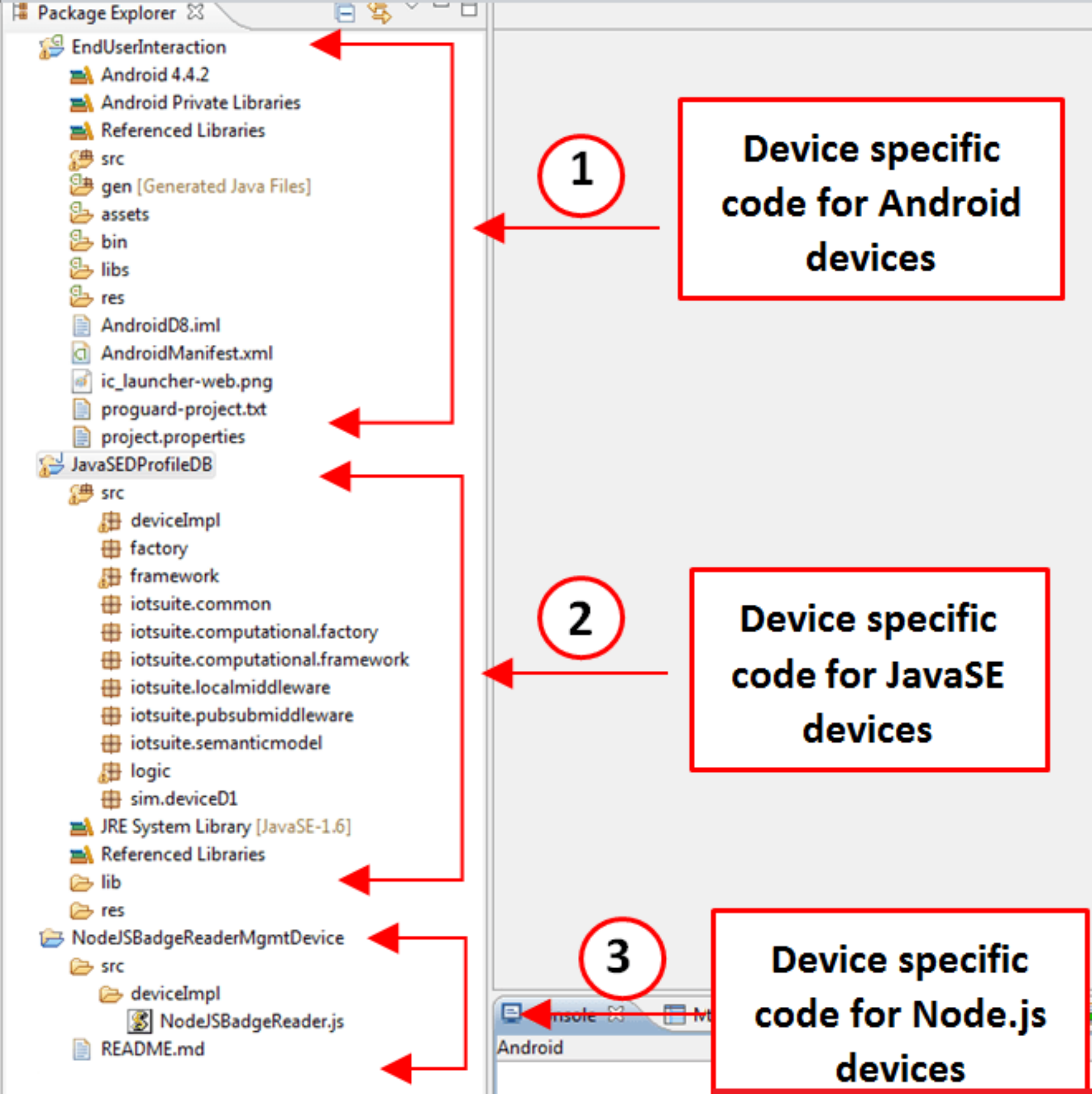} 
	\caption{Packages for target devices in Eclipse} \label{fig:eclipselinker}
\end{figure*}

The current version of the linker generates packages for Android, Node.js, and JavaSE platform. Figure~\ref{fig:eclipselinker} illustrates packages for Android devices~(\circled{1} in Figure~\ref{fig:eclipselinker}), JavaSE target devices~(\circled{2} in Figure~\ref{fig:eclipselinker}), and Node.js devices~(\circled{3} in Figure~\ref{fig:eclipselinker}) 
and  imported into Eclipse IDE. In order to execute code, these packages still need to be compiled by 
a device-level compiler designed for a target platform.
\subsection{Runtime system}\label{sec:runtimesys}
The main responsibility of the runtime system is 
a distributed execution of IoT applications~\cite{5283906}.  It is divided 
into three parts:  (1) \emph{middleware}: It runs on each individual device and provides 
a support for executing distributed tasks. (2) \emph{wrapper}: It plugs packages, generated by the linker module, and middleware. 
(3) \emph{support library}:  It separates packages,  produced by the linker component, and 
underlying middleware by providing interfaces that are implemented by each wrapper.
The integration of a new middleware into IoTSuite consists 
of an implementation of the following interfaces in the wrapper:

\fakeparagraph{\texttt{publish()}} It is an interface for publishing data from a sender.  The definition of this interface 
	    contains: an event name (e.g., temperature), event data (e.g., a temperature value, 
			 Celsius), and publisher's information such as location of a sender.
			
\fakeparagraph{\texttt{subscribe()}} It is an interface for receiving event notifications. An interest of events 
	is expressed by sending a subscription request, which contains: 
			  a event name~(e.g., temperature), information for filtering events such as regions 
				of interest~(e.g., a \texttt{RoomAvgTemp} component wants to receive events only from 
				a current room), and subscriber's information.
				
\fakeparagraph{\texttt{command()}} It is an interface for triggering an action of an actuator. A command contains: 
		a command name (e.g., switch-on \texttt{Heater}), command parameters~(e.g., set temperature of \texttt{Heater} to 30$^\circ$C), 
			 and a sender's information.
 
\fakeparagraph{\texttt{request-response()}} It is an interface for requesting data from a requester. In reply, a receiver 
		 sends a response. A request contains a request name (e.g., give profile information), request parameters~(e.g., give 
		  profile of  person with identification 12), and information about the requester.

The current implementation of IoTSuite uses the MQTT middleware. It enables interactions among Android devices, Node.js-enabled devices, and  JavaSE-enabled devices. The current wrapper implementation for the MQTT middleware is available at URL\footnote{\url{https://github.com/pankeshlinux/ToolSuite}}. 
\subsection{Eclipse plug-in}\label{sec:eclipseplugin}
We have integrated the above mentioned components as Eclipse plug-in to provide end-to-end support 
\begin{figure*}[!ht]
	\centering \includegraphics[width=0.7\linewidth]{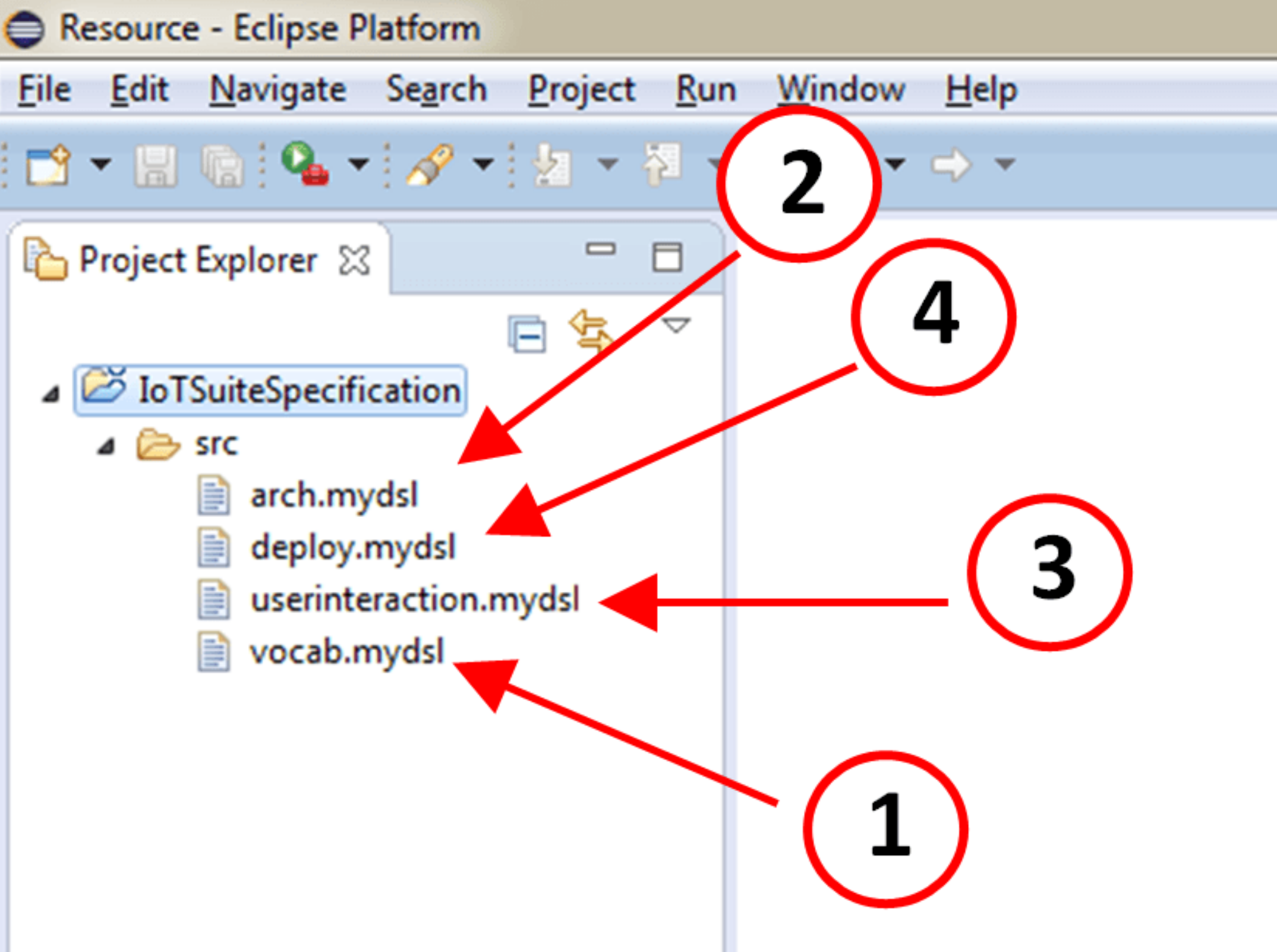} 
	\caption{Eclipse plug-in for IoT application development.} \label{fig:endtoend}
\end{figure*}
for IoT application development. Figure~\ref{fig:endtoend} illustrates use of our plug-in at  various phases of IoT application development:
\begin{enumerate}
	\item Vocab.mydsl~(\circled{1}  in Figure~\ref{fig:endtoend})  -- using which the domain expert can describe and compile a domain specification of an application domain. 				
 \item Arch.mydsl~(\circled{2} in Figure~\ref{fig:endtoend}) -- 
      using which the software designer can describe and compile an architecture specification of an application.  
			 \item Userinteraction.mydsl~(\circled{3} in  Figure~\ref{fig:endtoend}) -- 
     using which the software designer can describe and compile an userinteraction specification of an application.
 \item Deploy.mydsl~(\circled{4}  in Figure~\ref{fig:endtoend}) --  
         using which the network manager can describe a deployment specification of a target domain and invoke the mapping component. The network manager can combines and packs code generated by various stages 
		 of compilation into packages that can be deployed on devices.
  \end{enumerate}
\section{A step-by-step applications development using IoTSuite}\label{sec:appdev}

This section describes each step of IoT application development process using IoTSuite. Application development using
IoTSuite is a multi-step process and focuses on design, implement, and deployment
phases to develop IoT applications. 
We take different class of application~(SCC (Sense-Compute-Control), End-User-Interaction, and Data Visualization) as an example to demonstrate application development using IoTSuite.

\textit{Specifying high-level specifications.} It includes specifications of domain specification~(includes concepts that are specific to domain of IoT), architecture specification~(includes concepts that are specific to functionality of IoT), userinteraction specification~(defines what interaction are required by an application), and deployment specification~-(describes a device and its properties in a target deployment). Stakeholders specify high-level specifications using IoTSuite-Eclipse-Plugin\footnote{https://github.com/chauhansaurabhb/IoTSuite-Eclipse-Plugin}. 

\subsection{Personalized HVAC application:}
\begin{figure}[!ht]
	\centering 
	\includegraphics[width=1.0\linewidth]{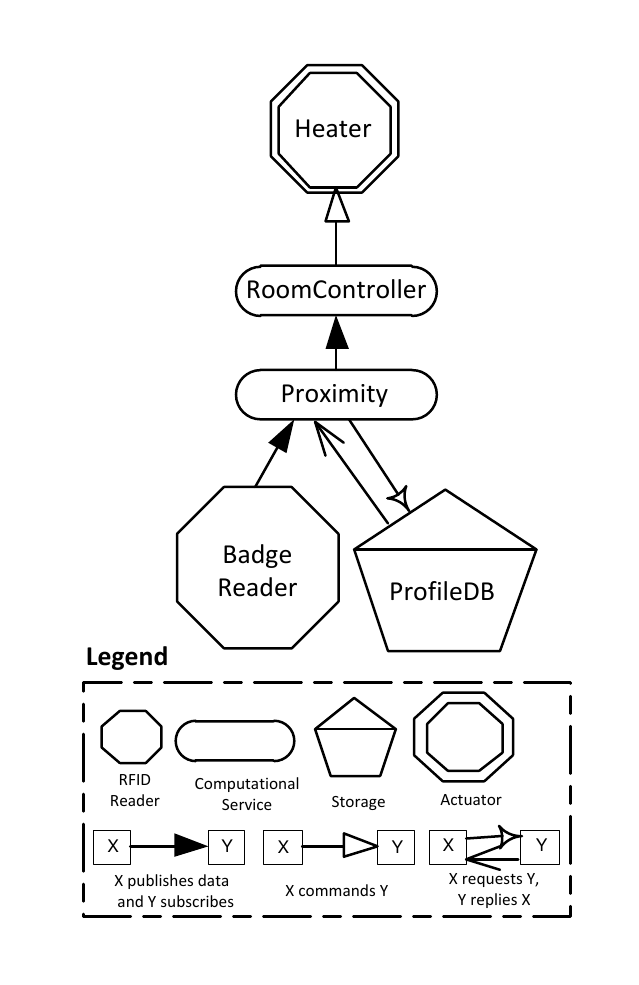} 
	\caption{Dataflow of Personalized-HVAC Application} \label{fig:phvac}
\end{figure}

A home consists of several rooms, each one is instrumented with several heterogeneous entities for providing resident's comfort. To accommodate a resident's preference in a room, a database is used to keep the profile of each resident, including his/her preferred temperature level. An RFID reader in the room detects the resident's entry and queries the database service. Based on this, the thresholds used by the room device~(Heater) are updated as shown in Figure~\ref{fig:phvac}. Developers need to follow following steps to develop above discussed application using IoTSuite.

\textit{Specifying Domain specification.} Developer specifies domain specification using IoTSuite as shown in Listing~\ref{vocabDSL}. Each resource is charaterized by types of information it generates or consumes. A set of information is defined using the \texttt{structs} keyword (Listing~\ref{vocabDSL}, line~\ref{vocab:structs-start}). For instance, a \texttt{BadgeReader} (lines~\ref{vocab:badgereader-tag-start}-\ref{vocab:badgereader-tag-end}) may generate a \texttt{badgeDetected} and \texttt{badgeDisappeared}. This information is defined as \texttt{BadgeStruct} and its two fields (\ref{vocab:badgestructs-start}-\ref{vocab:badgestructs-end}). A \texttt{Heater} is set according to a user's temperature preference illustrated in Listing~\ref{vocabDSL}, lines~\ref{vocab:actuator-heater-start}-\ref{vocab:actuators-end}. The \texttt{SetTemp} action takes a user's temperature preference shown in line~\ref{vocab:actuator-setTemp-start}. A set of storage is declared using the \texttt{storages}
keyword (Listing~\ref{vocabDSL}, line~\ref{vocab:storages-start}). A retrieval from the storage requires a parameter, specified using the \texttt{accessed-by} keyword (Listing~\ref{vocabDSL}, line~\ref{vocab:storages-generate-start}). For instance, a user's profile is accessed from storage by a \texttt{badgeID} (Listing~\ref{vocabDSL}, lines~\ref{vocab:storages-generate-start}-\ref{vocab:storages-generate-end}).

\lstset{emph={String, generate, actuators, structs, double,  action, resources, long, storages, accessed, by, sample, period, for, eventDrivenSensors, onCondition, true, tags, =}, emphstyle={\color{blue}\bfseries\emph}, caption={Code snippet of domain spec.}, escapechar=\#, 
 label=vocabDSL}	
\begin{lstlisting}
 structs:   #\label{vocab:structs-start}#
   TempStruct               #\label{vocab:tempstruct-start}#
    tempValue : double;
    unitOfMeasurement : String;	 #\label{vocab:structs-end}#
   BadgeStruct
      badgeID: String; #\label{vocab:badgestructs-start}#
      badgeEvent: String; #\label{vocab:badgestructs-end}#
  resources:
   tags:   #\label{vocab:tag-start}#
    BadgeReader             #\label{vocab:badgereader-tag-start}#     
      generate badgeDetected: BadgeStruct; 
	    generate badgeDisappeared: BadgeStruct;	#\label{vocab:badgereader-tag-end}# 	
  actuators:      #\label{vocab:actuators-start}#
    Heater   #\label{vocab:actuator-heater-start}#   
      action SetTemp(setTemp:TempStruct);	 #\label{vocab:actuator-setTemp-start}# 
	    action Off();    #\label{vocab:actuators-end}#
  storages:        #\label{vocab:storages-start}#
    ProfileDB  #\label{vocab:profile-start}#       
      generate profile: TempStruct accessed-by  #\label{vocab:storages-generate-start}#
					badgeID:String; #\label{vocab:storages-generate-end}#
\end{lstlisting}

\begin{figure*}[!ht]
	\centering 
	\includegraphics[width=0.55\linewidth]{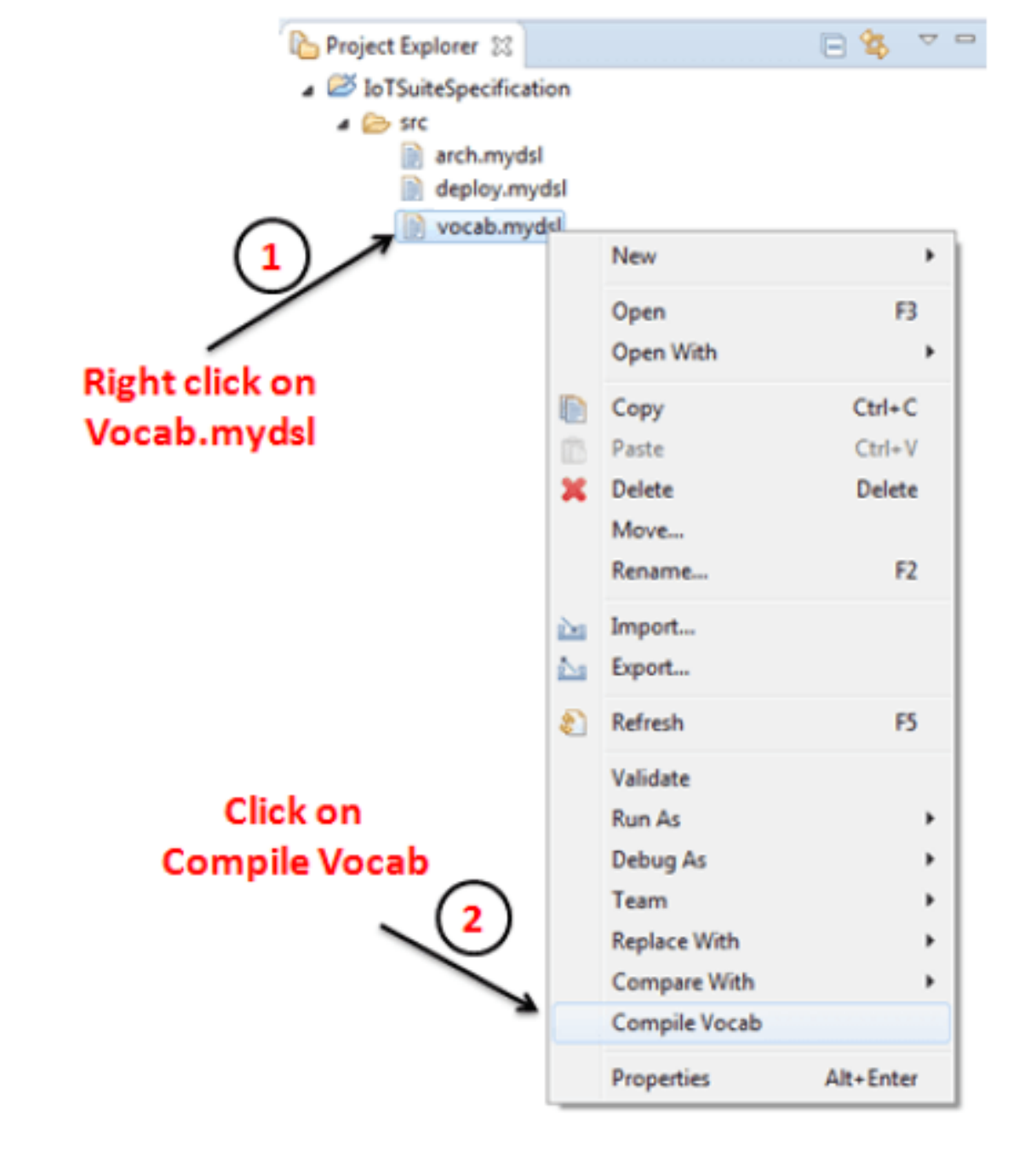} 
	\caption{Compilation of domain spec.} \label{fig:cvocab}
\end{figure*}

\textit{Compilation of domain specification.} To compile domain specification~(\texttt{vocab.mydsl} file), Right click on the \texttt{vocab.mydsl} file (Step \circled{1}) and click on Compile Vocab option (Step \circled{2}) as shown in Figure~\ref{fig:cvocab}.

\textit{Architecture specification.} Developer specifies architecture specification using IoTSuite as shown in Listing~\ref{archdsl}. It is described as a set of computational services. It consists
of two types of computational services: (1) \texttt{Common} component specifies common operations (e.g., average , count ,
sum ) in the application logic. (2) \texttt{Custom} specifies an application specific
logic. For instance, the \texttt{Proximity} component is a custom component that coordinates events from \texttt{BadgeReader} with the content from \texttt{ProfileDB} as shown in Figure~\ref{fig:phvac}. Each computational service is described by a set of inputs and outputs. For instance, the \texttt{Proximity} consumes \texttt{badgeDetected} and \texttt{badgeDisappeared} (Listing \ref{archdsl}, lines~\ref{arch-proximity:proximity-badgedetected-consume}-\ref{arch-proximity:proximity-badgedisappeared-consume}), request to \texttt{ProfileDB} to access user's profile (Listing \ref{archdsl}, line \ref{arch-proximity:proximity-request}) and generates \texttt{tempPref} (Listing \ref{archdsl}, line~\ref{arch-proximity:proximity-generate}).  Command is issued by a computational service to trigger an actions. For instance, the \texttt{RoomController} issues a \texttt{SetTemp} command (Listing \ref{archdsl}, line~\ref{arch-controller:controller-command}) with a \texttt{setTemp} as an argument to \texttt{Heater} (Listing~\ref{vocabDSL}, lines ~\ref{vocab:actuator-heater-start}-\ref{vocab:actuator-setTemp-start}).

\lstset{emph={generate, computationalServices, command, from, to, 
consume, request, Custom, Common, COMPUTE, AVG_BY_SAMPLE }, emphstyle={\color{blue}\bfseries\emph} }

\lstset{caption={A code snippet of architecture spec.}, escapechar=\#, label=archdsl}
\begin{lstlisting}
computationalServices:
 Custom: #\label{arch-custom-component}#
  Proximity #\label{arch-proximity:proximity-start}#
	   consume badgeDetected from BadgeReader;#\label{arch-proximity:proximity-badgedetected-consume}#
	   consume badgeDisappeared from BadgeReader;#\label{arch-proximity:proximity-badgedisappeared-consume}#
	   request profile to ProfileDB;#\label{arch-proximity:proximity-request}#
	   generate tempPref: TempStruct; #\label{arch-proximity:proximity-generate}#
  RoomController #\label{arch-controller:regulatetemp-start}#
	   consume tempPref from Proximity; #\label{arch-controller:proximity}#
	   command SetTemp(setTemp) to Heater;  #\label{arch-controller:controller-command}# 
	   command Off() to Heater;  #\label{arch-controller:controller-offcommand}# 
\end{lstlisting}

\begin{figure*}[!ht]
	\centering 
	\includegraphics[width=0.5\linewidth]{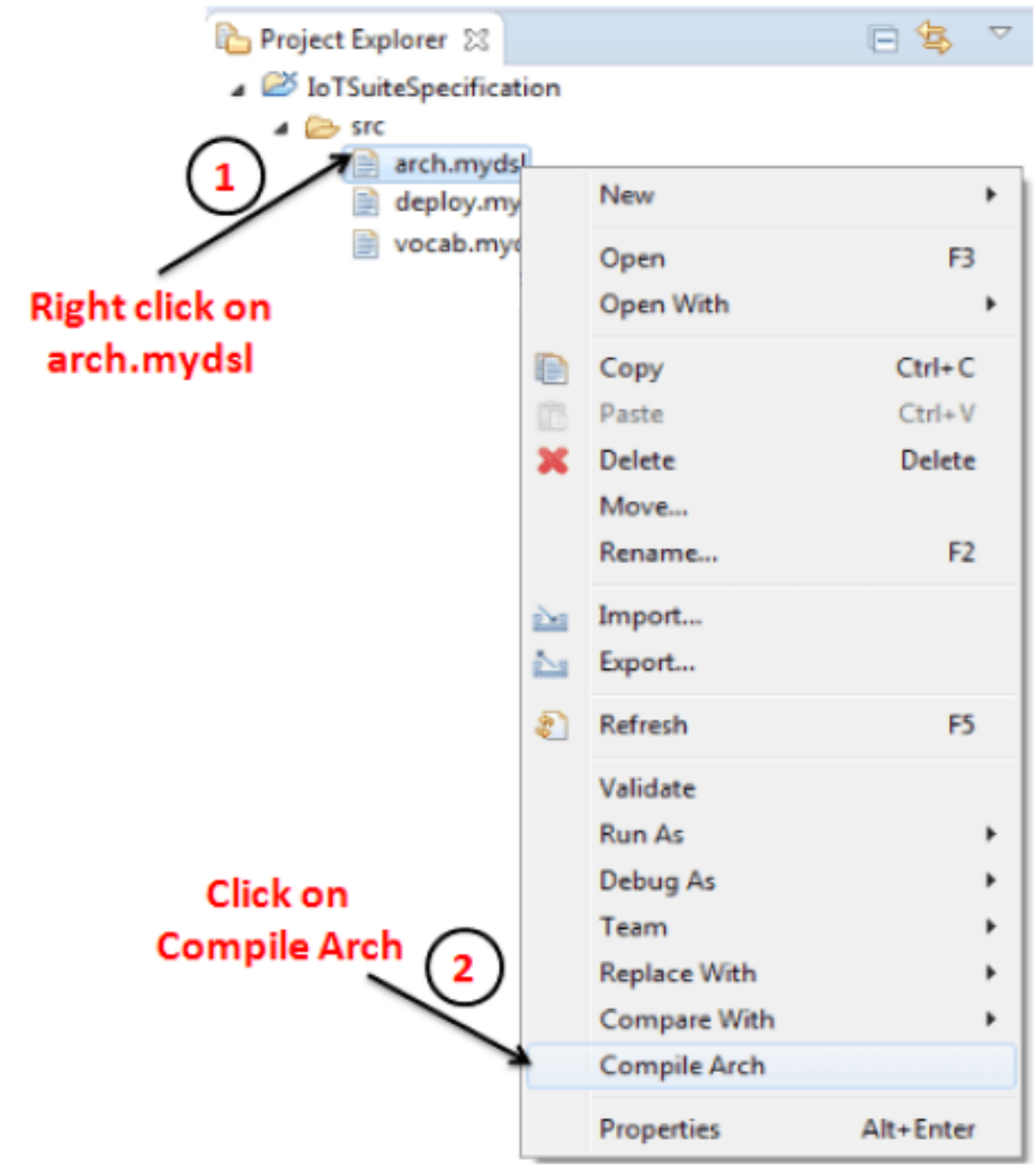} 
	\caption{Compilation of architecture spec.} \label{fig:carch}
\end{figure*}

\textit{Compilation of architecture specification.} To compile architecture specification~(\texttt{arch.mydsl} file), Right click on the \texttt{arch.mydsl} file (Step \circled{1}) and click on Compile Arch option (Step \circled{2}) as shown in Figure~\ref{fig:carch}.

\begin{figure*}[!ht]
\centering
\includegraphics[width=0.6\textwidth]{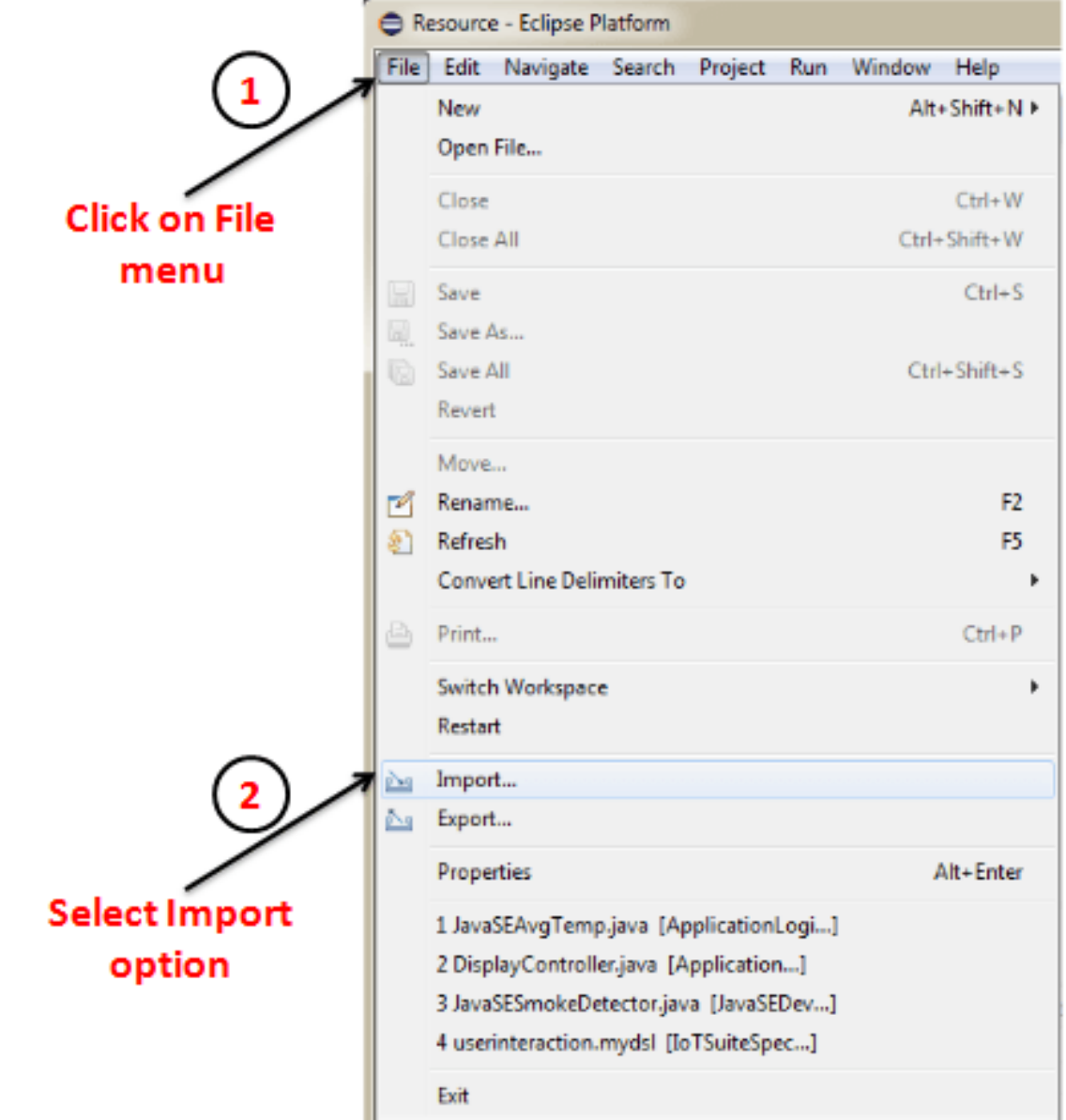}
\caption{Import application logic package }
\label{fig:importlogic}
\end{figure*}

\textit{Import application logic package.} To import application logic package, click on File Menu (Step \circled{1}), and select Import option (Step \circled{2}) as shown in Figure~\ref{fig:importlogic}.

\begin{figure*}[!ht]
\centering
\includegraphics[width=0.9\textwidth]{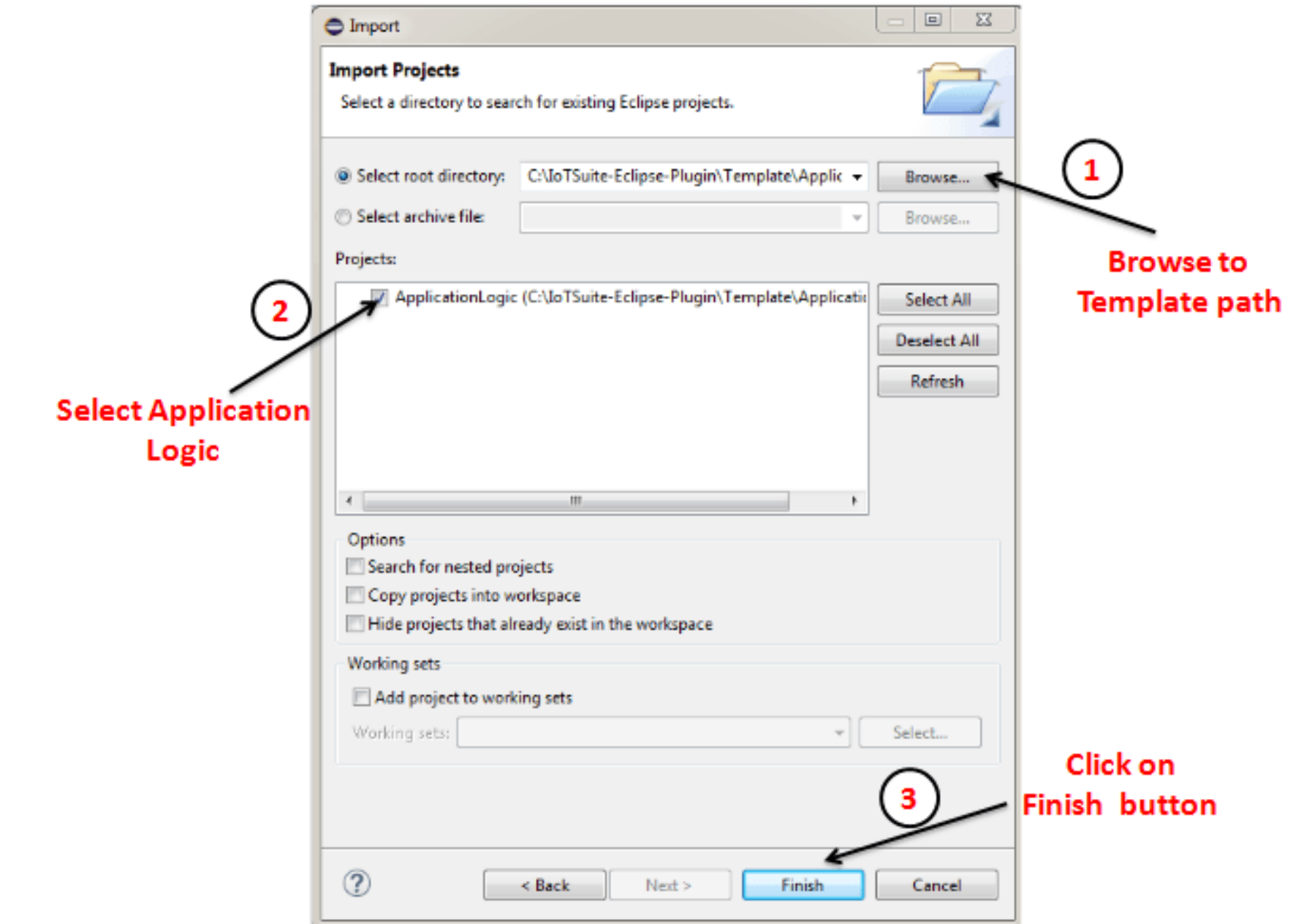}
\caption{Locate application logic package }
\label{fig:locatelogic}
\end{figure*}

\textit{Locate application logic package.} To locate application logic package, browse to Template path (Step \circled{1}), select application logic package (Step \circled{2}), and click on Finish button (Step \circled{3}) as shown in Figure~\ref{fig:locatelogic}.

\textit{Implement application logic.} The application logic proj-ect contains a generated framework that hides low-level details from a developer and allows the developer to focus on the application logic. The developer writes application
specific logic in logic package. Listing \ref{proximityclass} and \ref{roomcontrollerclass} show generated Java programming framework from the architecture specification, defined in Listing~\ref{archdsl}. To implement application logic of \texttt{Proximity} component, developers have to implement the generated abstract methods illustrated in Listing \ref{proximityclass}. The \texttt{Proximity} component receives the \texttt{badgeDetected} and \texttt{badgeDisappeared} events from the \texttt{BadgeReader} and coordinates with \texttt{ProfileDB} component for user-temperature preference defined in Listing \ref{archdsl}, lines~\ref{arch-proximity:proximity-start}-\ref{arch-proximity:proximity-request}. To implement this logic, three methods (\texttt{onNewbadgeDetected}, \texttt{onNewbadgeDisappeared}, and \texttt{onNewprofileReceived}) have to implemented as shown in Listing~\ref{proximityclass}. It request to \texttt{ProfileDB} for user's preference using \texttt{onNewbadgeDetected} method, turn off \texttt{Heater} using \texttt{onNewbadgeDisappeared} method, and generate \texttt{tempPref} using \texttt{onNewprofileReceived} method which is consumed by the \texttt{RoomController} defined in Listing~\ref{archdsl}, line~\ref{arch-controller:proximity}.  In similar ways, developer implements application logic of \texttt{RoomController} component using \texttt{onNewtempPref} method. It issues a \texttt{SetTemp} and \texttt{Off} commands to \texttt{Heater} using \texttt{onNewtempPref} method as illustrated in Listing~\ref{roomcontrollerclass}.

\lstset{language=Java, caption={The implementation of the Java
    abstract class \texttt{Proximity} written by the developer.}, escapechar=\#, label=proximityclass}
\begin{lstlisting}
package logic;
import android.content.Context;
import iotsuite.semanticmodel.*;
import framework.*;

public class LogicProximity extends Proximity {

	public LogicProximity(PubSubMiddleware pubSubM, 
			Device deviceInfo, Object ui, Context myContext) {
		super(pubSubM, deviceInfo);
	}
	
	@Override
	public void onNewbadgeDetected(BadgeStruct arg) {
		// Request profileDB for his preference
		getprofile(arg.getbadgeID());
	}	
	@Override
	public void onNewbadgeDisappeared(BadgeStruct arg) {
		// Person is leaving the room. Turn off the heater.
		TempStruct tempStruct = new TempStruct(-100, "C");
		settempPref(tempStruct);
	}	
	public void onNewprofileReceived(TempStruct arg) {
		TempStruct tempStruct = new TempStruct(arg.gettempValue(), "C");
		// Set temperature to RoomController
		settempPref(tempStruct);
	}
}
\end{lstlisting}

\lstset{language=Java, caption={The implementation of the Java
    abstract class \texttt{RoomController} written by the developer.}, escapechar=\#, label=roomcontrollerclass}
\begin{lstlisting}
package logic;

import iotsuite.pubsubmiddleware.PubSubMiddleware;
import android.content.Context;
import iotsuite.semanticmodel.*;
import framework.*;

public class LogicRoomController extends RoomController {

	public LogicRoomController(PubSubMiddleware pubSubM, 
		   Device deviceInfo, Object ui, Context myContext) {
		super(pubSubM, deviceInfo);
	}

	@Override
	public void onNewtempPref(TempStruct arg) {
		if (arg.gettempValue() == -100) { 
		// It means that person is leaving the room
			Off(); // Set the Off() command
		} else {
			// If the person is entering to the room then, 
			// set temperature according to his preference
			double tempValue = arg.gettempValue();
			TempStruct tempStruct = new TempStruct(tempValue,
					arg.getunitOfMeasurement());
			SetTemp(tempStruct);
		}
	}
}
\end{lstlisting}

\textit{Deployment specification.} Developer specifies deployment specification using IoTSuite as shown in Listing~\ref{deploydsl}.  It includes properties such as \texttt{location} that defines where a device is deployed, \texttt{resources} define component(s) to be deployed on a device, \texttt{language-platform} is used to generate an appropriate pack-age for a device, \texttt{protocol} specifies a run-time system in-stalled on a device to interact with other devices. Listing~\ref{deploydsl} shows a code snippet to illustrate these concepts. \texttt{BadgeReaderMgmtDevice} is located in room\#1 (line \ref{BadgeReader-location}),  \texttt{BadgeReader} is attached with the device (line \ref{BadgeReader-resources}) and the device driver code for these two component is in \texttt{NodeJS} (line \ref{BadgeReader-platform}), \texttt{mqtt} runtime system is installed on \texttt{BadgeReaderMgmtDevice} (line \ref{BadgeReader-protocol}). A storage device contains the \texttt{database} field that specifies the installed database. This field is used to select an appropriate storage driver. For instance, \texttt{DatabaseSrvDevice} (line \ref{Database-component}) runs \texttt{ProfileDB} (line \ref{Database-resources}) component implemented in \texttt{MySQL} database (line \ref{Database-database}) as illustrated in Listing \ref{deploydsl}.

\lstset{emph={resources, softwarecomponents,  devices, location, protocol, description,database, language,platform}, emphstyle={\color{blue}\bfseries\emph}%
}

\lstset{caption={Code snippet of deployment spec.}, label=deploydsl}
\begin{lstlisting}
devices:												
	BadgeReaderMgmtDevice:    #\label{BadgeReader-component}#                                            
		    location:                                       
			  Room:1;               #\label{BadgeReader-location}#                             
			platform: NodeJS;     #\label{BadgeReader-platform}#                             
			resources: BadgeReader;  #\label{BadgeReader-resources}#                    
			protocol: mqtt;    #\label{BadgeReader-protocol}#                               
	DatabaseSrvDevice:         #\label{Database-component}#                                          
			location:                                       
	 		  Room: 1;                                  
	 		platform: JavaSE;                           
	 		resources: ProfileDB;   #\label{Database-resources}#                     
	 		protocol: mqtt;                          
	 		database: MySQL;  #\label{Database-database}#
	ResourceMgmtDevice1:                                                
			location:                                       
	 		  Room: 1;                                  
	 		platform: JavaSE;                           
	 		resources: ;                      
	 		protocol: mqtt; 
	ResourceMgmtDevice2:                                                
			location:                                       
	 		  Room: 1;                                  
	 		platform: JavaSE;                           
	 		resources: ;                      
	 		protocol: mqtt;    		
	HeaterMgmtDevice:                                                
			location:                                       
	 		  Room: 1;                                  
	 		platform: JavaSE;                           
	 		resources: Heater;                      
	 		protocol: mqtt;                          
	 		
	                      
	 		                           
\end{lstlisting}

\begin{figure*}[!ht]
\centering
\includegraphics[width=0.6\textwidth]{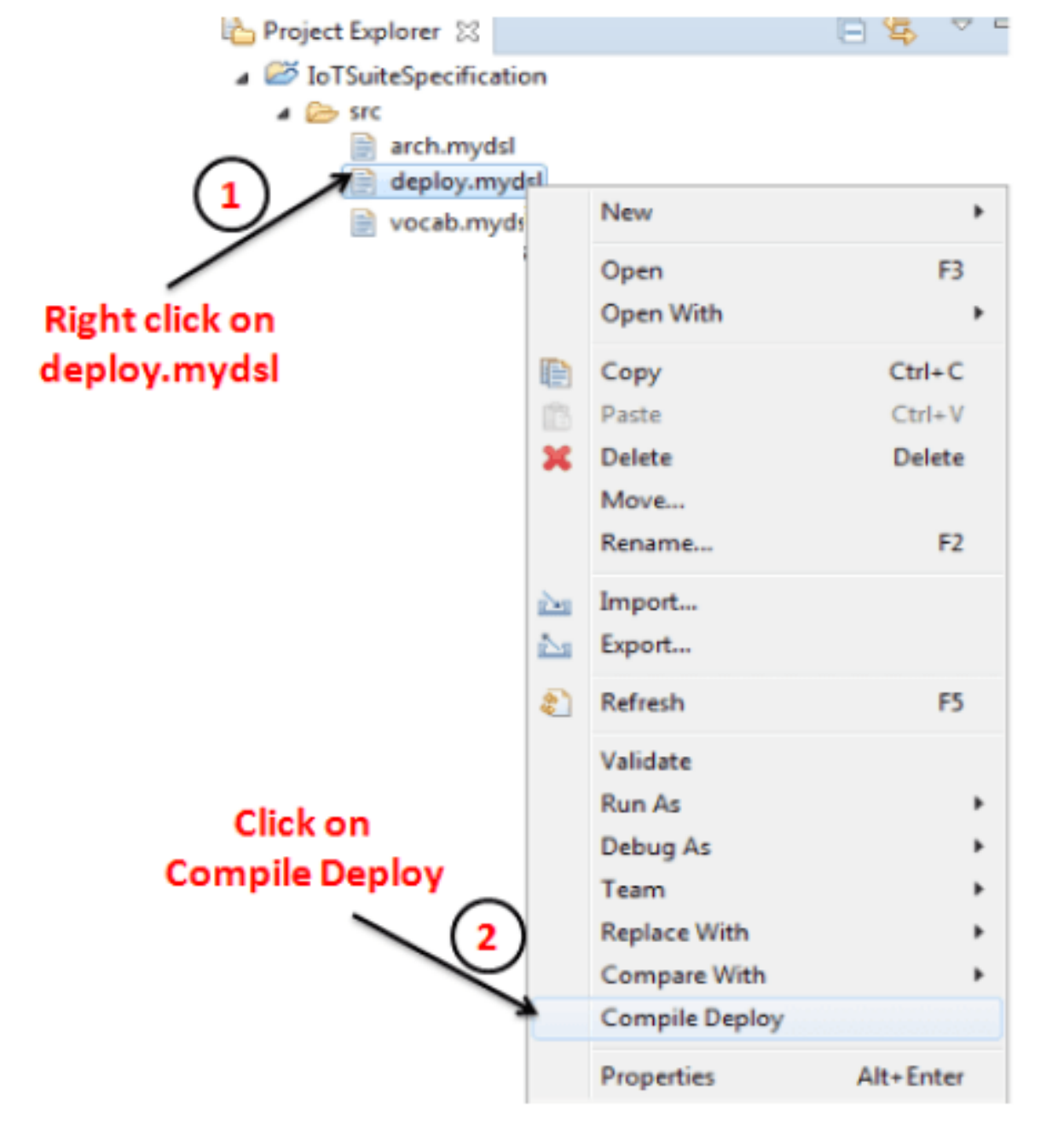}
\caption{Compilation of deployment spec.}
\label{fig:cdeploy}
\end{figure*}

\textit{Compilation of deployment spec.} Right click on \texttt{deploy.mydsl} file (Step \circled{1}) and selecting Compile Deploy (Step \circled{2}) as shown in Figure~\ref{fig:cdeploy} generates a deployment packages.

\begin{figure*}[!ht]
\centering
\includegraphics[width=0.8\textwidth]{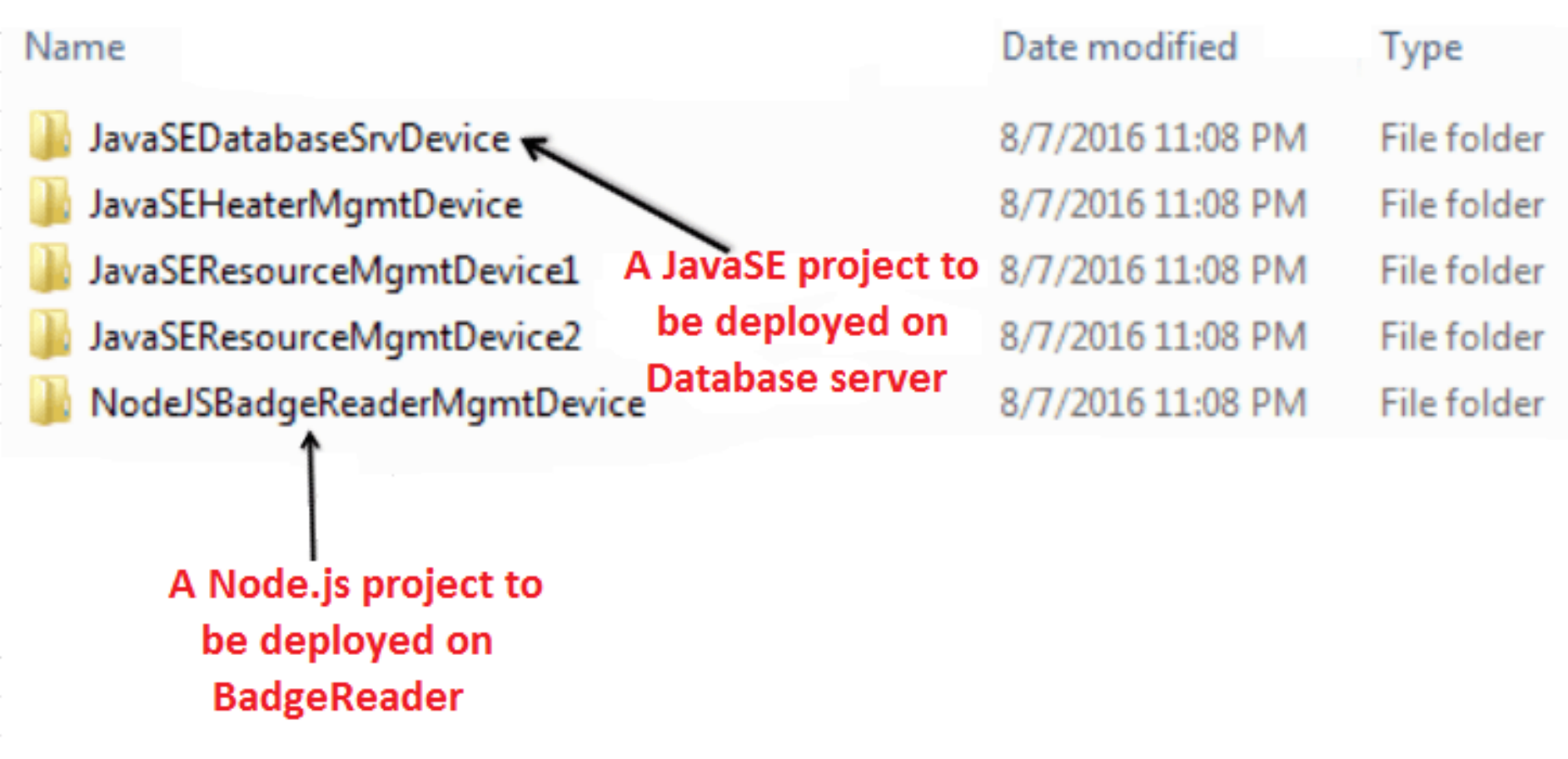}
\caption{Packages for target devices specified in the deployment spec.}
\label{fig:deploypackage}
\end{figure*}

\textit{Deployment of generated packages.} The output of compilation of deployment specification produce a set of platform
specific project/packages as shown in Figure~\ref{fig:deploypackage} for devices, specified in the
deployment specification (Refer Listing~\ref{deploydsl}). These projects compiled by device specific compiler designed for the target platform. The generated packages integrate the run-time system.

\subsection{Fire Detection Application}
\begin{figure}[!ht]
	\centering 
	\includegraphics[width=1.1\linewidth]{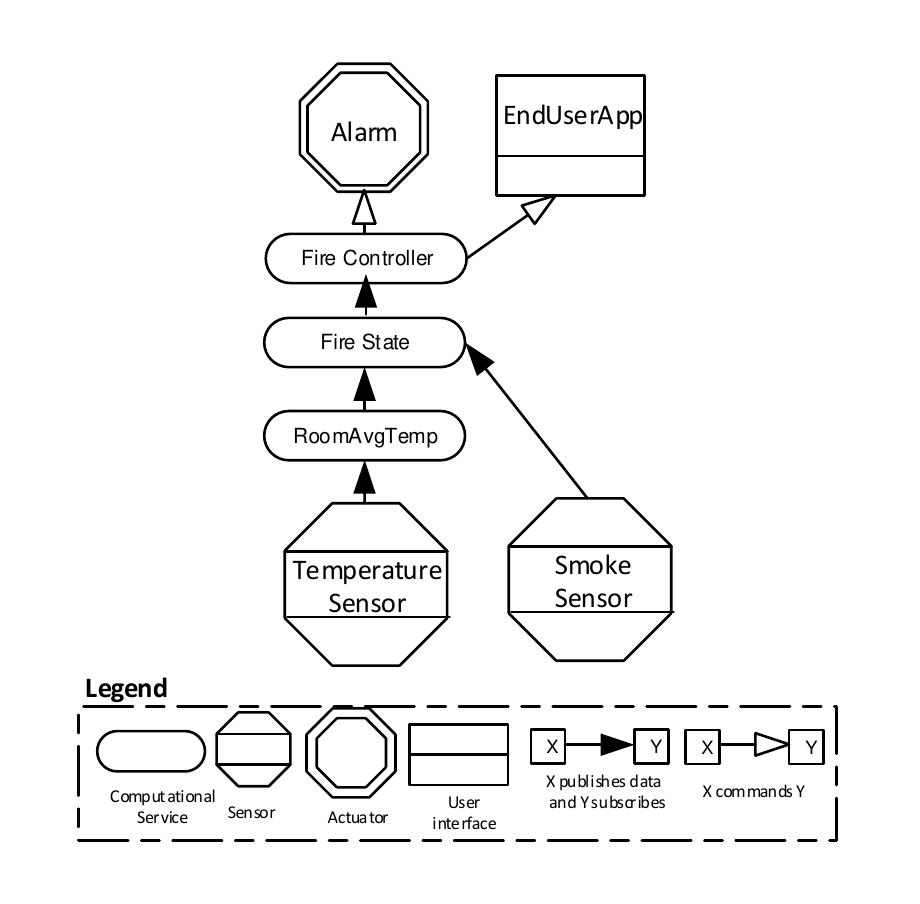} 
	\caption{Dataflow of Fire Detection Application} \label{fig:eu}
	\end{figure}

A home consists of several rooms, each one is instrumented with several heterogeneous entities for providing resident's safety.
To ensure the safety of residents, a fire detection application is installed. It aims to detect fire by analyzing data from smoke 
 and temperature sensors. When fire occurs, residences are notified on their smart phones by an installed application. Additionally, residents and their neighbors are informed through a set of alarms as shown in Figure~\ref{fig:eu}.
	
\textit{Specifying Domain specification.} Developer specifies domain specification using IoTSuite as shown in Listing~\ref{euvocabDSL}. Each resource is charaterized by types of information it generates or consumes. A set of information is defined using the \texttt{structs} keyword (Listing~\ref{euvocabDSL}, line~\ref{vocab:structs-eustart}). A set of sensors is declared using the \texttt{sensors} keyword (Listing \ref{euvocabDSL}, line \ref{vocab:sensors-eustart}). Each sensor produces one or more sensor measurements along with the data-types specified in the data structure (Listing \ref{euvocabDSL}, lines \ref{vocab:tsstructs-eustart}-\ref{vocab:tsstructs-euend}), declared using the \texttt{generate} keyword. A periodic sensor samples results every d seconds for duration k seconds. For instance, a temperature sensor generates a \texttt{tempMeasurement} (Listing\ref{euvocabDSL}, line \ref{vocab:eugenerate}) of \texttt{TempStruct} type (Listing \ref{euvocabDSL}, lines \ref{vocab:tsstructs-eustart}-\ref{vocab:tempstructs-euend}). It samples data every 1 second for next 6 minutes (Listing \ref{euvocabDSL}, line \ref{vocab:eusample}). An event-driven sensor produces data when the event condition is met. For instance, a smoke sensor generates \texttt{smokeMeasurement} when \texttt{smokeValue > 650}
\texttt{PPM} (Listing \ref{euvocabDSL}, lines \ref{vocab:eusmoke-start}-\ref{vocab:eusmoke-end}). A set of actuators is declared using the \texttt{actuators} keyword (Listing \ref{euvocabDSL}, line \ref{vocab:actuator-eustart}). Each actuator has one or more actions declared using the \texttt{action} keyword. For instance, an \texttt{Alarm} may have one action (e.g, on), illustrated in Listing \ref{euvocabDSL}, line \ref{vocab:actuator-euend}.

\lstset{emph={String, generate, actuators, structs, double,  action, resources, long, storages, accessed, by, sample, period, for, eventDrivenSensors, periodicSensors, onCondition, true, tags, =}, emphstyle={\color{blue}\bfseries\emph}, caption={Code snippet of domain spec.}, escapechar=\#, 
 label=euvocabDSL}	
\begin{lstlisting}
structs:				#\label{vocab:structs-eustart}#																	
 TempStruct            #\label{vocab:tsstructs-eustart}#                                                                 
	tempValue: double;                                                                  
	unitOfMeasurement : String;    #\label{vocab:tempstructs-euend}# 
 SmokeStruct                                                                             
	smokeValue:double;                                                                  
	unitOfMeasurement:String;  #\label{vocab:tsstructs-euend}#                                                            
resources:                                                                                 
 sensors:     #\label{vocab:sensors-eustart}#	
  periodicSensors:                                                                    
	TemperatureSensor                                                               
	 generate tempMeasurement:TempStruct;    #\label{vocab:eugenerate}#                                    
	 sample period 1000 for 6000000;      #\label{vocab:eusample}#                                        
  eventDrivenSensors:                                                                 
	SmokeDetector                                                                   
	 generate smokeMeasurement:SmokeStruct;  #\label{vocab:eusmoke-start}#                                       
	 onCondition smokeValue > 650 PPM ;  #\label{vocab:eusmoke-end}#  
  actuators:     #\label{vocab:actuator-eustart}#                                                                       
	Alarm                                                                           
	 action On();   #\label{vocab:actuator-euend}#                 
					
\end{lstlisting}

\textit{Compilation of domain specification.} To compile domain specification~(\texttt{vocab.mydsl} file), Right click on the \texttt{vocab.mydsl} file (Step \circled{1}) and click on Compile Vocab option (Step \circled{2}) as shown in Figure~\ref{fig:cvocab}.

\textit{Architecture specification.} Developer specifies architecture specification using IoTSuite as shown in Listing~\ref{euarchdsl}. It is described as a set of computational services. It consists of two types of computational services: (1) \texttt{Common} component specifies common operations (e.g., average, count, sum ) in the application logic. For instance, \texttt{RoomAvgTemp} component consumes 5 temperature measurements (Listing \ref{euarchdsl}, lines \ref{arch-avg-consume}-\ref{arch-avg-start}), apply average by sample operation (Listing \ref{euarchdsl}, line \ref{arch-avg-start}), and generates \texttt{roomAvgTempMeasurement} (Listing \ref{euarchdsl}, line \ref{arch-avg-generate}). (2) \texttt{Custom} component specifies an application specific logic. Additionally, each computational service is described by a set of inputs and outputs. For instance, the \texttt{FireController} consumes \texttt{smokeValue} (Listing \ref{euarchdsl}, line~\ref{arch-euconsume-start}), issues On and FireNotify (with \texttt{fireNotify} as argument) commands (Lines~\ref{arch-command-start}-\ref{arch-eucommand-end}). 


\lstset{emph={generate, computationalServices, command, from, to, 
consume, request, Custom, Common, COMPUTE, AVG_BY_SAMPLE }, emphstyle={\color{blue}\bfseries\emph} }

\lstset{caption={A code snippet of architecture spec.}, escapechar=\#, label=euarchdsl}
\begin{lstlisting}
computationalService:                                                       
 Common:                                                                 
  RoomAvgTemp                                                             
   consume tempMeasurement from TemperatureSensor;   #\label{arch-avg-consume}#               
   COMPUTE (AVG_BY_SAMPLE,5);     #\label{arch-avg-start}#                               
   generate roomAvgTempMeasurement :TempStruct;  #\label{arch-avg-generate}#   
 Custom:                    
  FireState                                                          
   consume roomAvgTempMeasurement from RoomAvgTemp;     #\label{fire-consume-start}#               
   consume smokeMeasurement from SmokeDetector; #\label{fire-consume-end}#
   generate smokeValue:SmokeStruct;       #\label{fire-generate}#                      
  FireController                                                     
   consume smokeValue from FireState; #\label{arch-euconsume-start}#                             
   command On() to Alarm;               #\label{arch-command-start}#                           
   command FireNotify(fireNotify) to EndUserApp; #\label{arch-eucommand-end}# 	 	    
			
\end{lstlisting}

\textit{Compilation of architecture specification.} To compile architecture specification~(\texttt{arch.mydsl} file), Right click on the \texttt{arch.mydsl} file (Step \circled{1}) and click on Compile Arch option (Step \circled{2}) as shown in Figure~\ref{fig:carch}.
%

\textit{Import application logic package.} To import application logic package, click on File Menu (Step \circled{1}), and select Import option (Step \circled{2}) as shown in Figure~\ref{fig:importlogic}.


\textit{Locate application logic package.} To locate application logic package, browse to Template path (Step \circled{1}), select application logic package (Step \circled{2}), and click on Finish button (Step \circled{3}) as shown in Figure~\ref{fig:locatelogic}.

\textit{Implement application logic.} The application logic proj-ect contains a generated framework that hides low-level details from a developer and allows the developer to focus on the application logic. The developer writes application
specific logic in logic package. Listing \ref{firestate} and \ref{firecontroller} show generated Java programming framework from the architecture specification, defined in Listing~\ref{euarchdsl}. To implement application logic of \texttt{FireState}, developers have to implement the generated abstract methods illustrated in Listing \ref{firestate}. The \texttt{FireState} component consumes the \texttt{roomAvgTempMeasurement} from \texttt{RoomAvgTemp}, \texttt{smokeMeasurement} from \texttt{SmokeDetector}  and generates \texttt{smokeValue}  defined in Listing~\ref{euarchdsl}, lines \ref{fire-consume-start}-\ref{fire-generate}. To implement this logic, two methods \texttt{onNewsmokeMeasurement}, and \texttt{onNewroomAvgTempMeasurement} have to implemented as shown in Listing~\ref{firestate}. It reads the \texttt{smokeMeasurement} using \texttt{onNewsmokeMeasurement} m-ethod, \texttt{roomAvgTempMeasurement} using \texttt{onNewroomAvgTe-mpMeasurement} method, and setsmokeValue if smokeValue and avgtempValue are greater than threshold value. In similar ways, developer implements application logic of \texttt{FireController} component using \texttt{onNewtempPref} method. It issues On command to alarm and fireNotify command to EndUserApp using \texttt{onNewsmokeValue} method as illustrated in Listing~\ref{firecontroller} defined in Listing \ref{euarchdsl} lines \ref{arch-command-start}-\ref{arch-eucommand-end}.

%
%
%
\lstset{language=Java, caption={The implementation of the Java
    abstract class \texttt{FireState} written by the developer.}, escapechar=\#, label=firestate}
\begin{lstlisting}
package logic;

import iotsuite.pubsubmiddleware.PubSubMiddleware;
import android.content.Context;
import iotsuite.semanticmodel.*;
import framework.*;

public class LogicFireState extends FireState {
	double avgtempValue, smokeMeasurement;
	public LogicFireState(PubSubMiddleware pubSubM, Device deviceInfo,
			Object ui, Context myContext) {
		super(pubSubM, deviceInfo);
	}
	@Override
	public void onNewsmokeMeasurement(SmokeStruct arg) {
		// Read smokeMeasurement from SmokeDetector
		smokeMeasurement = arg.getsmokeValue();
		// If smokeValue and avgtempValue are greater than threshold means fire
		// has been occurred.
		if (smokeMeasurement > 650 && avgtempValue > 55) {
			SmokeStruct smokeStruct = new SmokeStruct(650, "PPM");
			setsmokeValue(smokeStruct);
		}
	}

	@Override
	public void onNewroomAvgTempMeasurement(TempStruct arg) {
		// Read roomAvgTempMeasurement from RoomAvgTemp
		avgtempValue = arg.gettempValue();
	}
}
\end{lstlisting}

\lstset{language=Java, caption={The implementation of the Java
    abstract class \texttt{FireController} written by the developer.}, escapechar=\#, label=firecontroller}
\begin{lstlisting}
package logic;

import java.sql.Timestamp;
import iotsuite.pubsubmiddleware.PubSubMiddleware;
import android.content.Context;
import iotsuite.semanticmodel.*;
import framework.*;

public class LogicFireController extends FireController {
	public LogicFireController(PubSubMiddleware pubSubM, Device deviceInfo,
			Object ui, Context myContext) {
		super(pubSubM, deviceInfo);
	}

	@Override
	public void onNewsmokeValue(SmokeStruct arg) {
		// Read smokeValue from FireState and fire command to Alarm and EndUserApp
		On();
		FireNotify(new FireStateStruct("Fire has been occurred",
				(new Timestamp(new java.util.Date().getTime())).toString()));
	}
}
\end{lstlisting}

\textit{Userinteraction specification.} Developer specifies userinteraction specification using IoTSuite as shown in Listing~\ref{euuidsl}. It defines what interactions are required by an application. We design a set of abstract interactors that denotes data exchange between an application and a user. \texttt{Notify} denotes information flow from an application to a user. For instance, an application notifies a user in case of fire. It is declared using the \texttt{notify} keyword (Listing \ref{euuidsl}, line \ref{eu-EndUserApp-start}). The application notifies users with the fire information specified in the data structure (Listing \ref{euuidsl}, lines \ref{eu-struct-start}-\ref{eu-struct-end}).

\lstset{emph={structs, resources,  userInteractions}, emphstyle={\color{blue}\bfseries\emph}%
}
\lstset{caption={Code snippet of user-interaction spec.}, label=euuidsl}
\begin{lstlisting}
structs:																									
 FireStateStruct   #\label{eu-struct-start}#                                                                                        
  fireValue:String;                                                                                   
  timeStamp:String;     #\label{eu-struct-end}#                                                                               
resources:                                                                                                 
 userInteractions:                                                                                       
  EndUserApp                                                                                          
  notify FireNotify(fireNotify:FireStateStruct) from FireController;  #\label{eu-EndUserApp-start}#                           
		
\end{lstlisting}

\begin{figure*}[!ht]
	\centering 
	\includegraphics[width=0.55\linewidth]{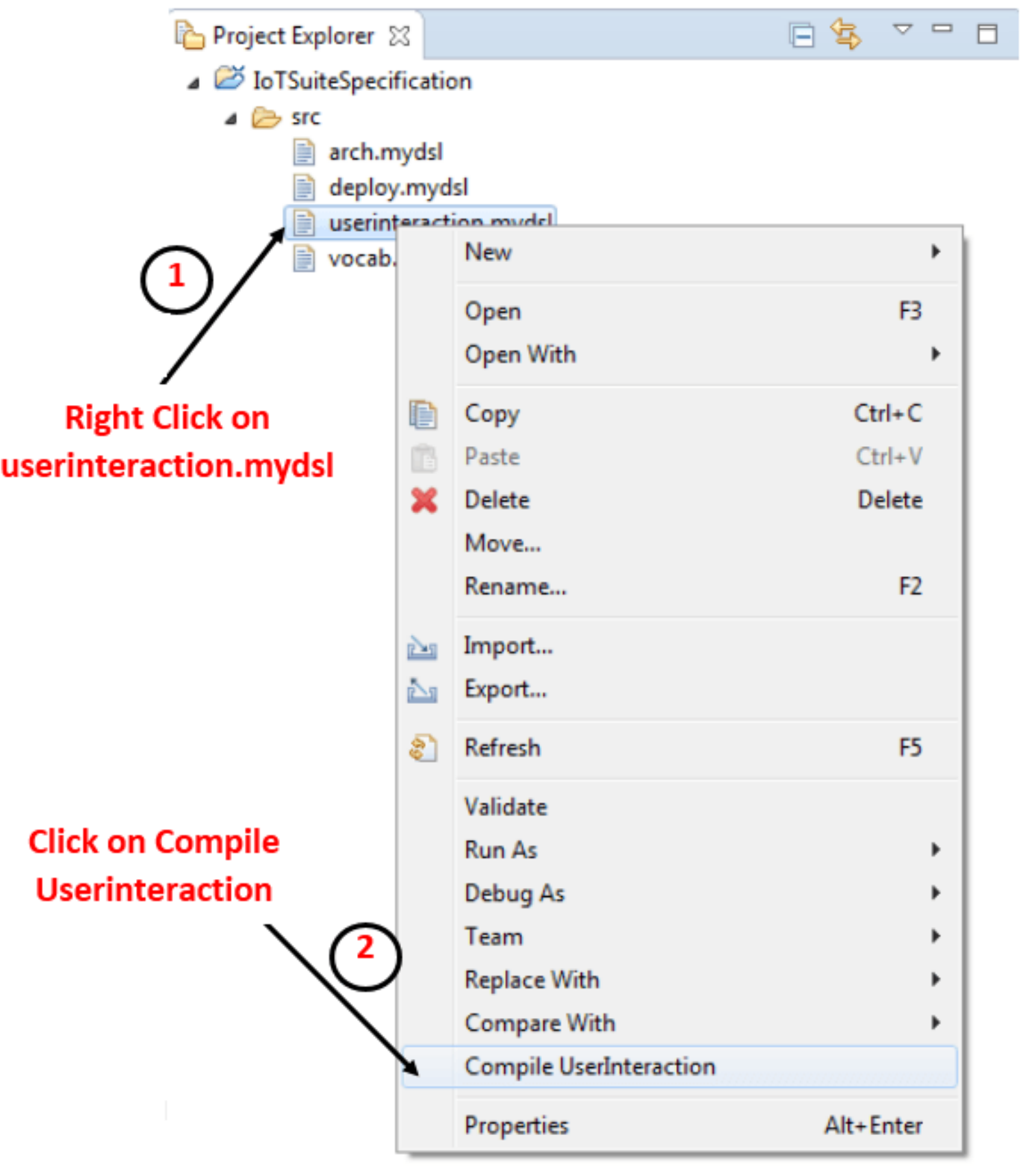} 
	\caption{Compilation of userinteraction spec.} \label{fig:eucui}
\end{figure*}

\textit{Compilation of userinteraction specification.} To compile userinteraction specification~(\texttt{userinteraction.mydsl} fil-e), Right click on the \texttt{userinteraction.mydsl} file (Step \circled{1}) and click on Compile UserInteraction option (Step \circled{2}) as shown in Figure~\ref{fig:eucui}.

\textit{Deployment specification.} Developer specifies deployment specification using IoTSuite as shown in Listing~\ref{eudeploydsl}. It includes properties such as \texttt{location} that defines where a device is deployed, \texttt{resources} define component(s) to be deployed on a device, \texttt{language-platform} is used to generate an appropriate package for a device, \texttt{protocol} specifies a run-time system in-stalled on a device to interact with other devices. Listing~\ref{eudeploydsl} shows a code snippet to illustrate these concepts. 
\texttt{SensorMgmtDevice1} is located in room\#1 (line \ref{SensorMgmtDevice-eu-room}),  \texttt{TemperatureSensor} is attached with the device (line \ref{SensorMgmtDevice-eu-resources}) and the device driver code for this component is in \texttt{NodeJS} (line \ref{SensorMgmtDevice-eu-platform}), \texttt{mqtt} runtime system is installed on \texttt{SensorMgmtDevice1} (line \ref{SensorMgmtDevice-eu-protocol}). 

\lstset{emph={resources, softwarecomponents,  devices, location, protocol, description,database, language,platform}, emphstyle={\color{blue}\bfseries\emph}%
}
\lstset{caption={Code snippet of deployment spec.}, label=eudeploydsl}
\begin{lstlisting}
devices:												
 SensorMgmtDevice1:        #\label{SensorMgmtDevice-eu-component}#                                             
  location:                                       
	 Room:1 ;                   #\label{SensorMgmtDevice-eu-room}#                   
  platform: NodeJS;                        #\label{SensorMgmtDevice-eu-platform}#    
  resources: TemperatureSensor;  #\label{SensorMgmtDevice-eu-resources}# 
  protocol: mqtt;                  #\label{SensorMgmtDevice-eu-protocol}#  
 SensorMgmtDevice2:                                                   
  location:                                       
	 Room:1 ;                                    
  platform: NodeJS;                       
  resources: SmokeDetector;  
  protocol: mqtt; 	
 ResourceMgmtDevice1:                                                
  location:                                       
	 Room: 1 ;                                  
  platform: JavaSE;                           
  resources: ;                      
  protocol: mqtt; 
 ResourceMgmtDevice2:                                                
  location:                                       
	 Room: 1 ;                                  
  platform: JavaSE;                           
  resources: ;                      
  protocol: mqtt;
 ResourceMgmtDevice3:                                                
  location:                                       
	 Room: 1 ;                                  
  platform: JavaSE;                           
  resources: ;                      
  protocol: mqtt;	
 EndUserDevice:                                                 
  location:                                       
	 Room:1 ;                                   
  platform: Android;         #\label{SensorMgmtDevice-Android-platform}#                  
  resources: EndUserApp;              
  protocol: mqtt; 		                       
	 		                           
\end{lstlisting}

\textit{Compilation of deployment spec.} Right click on \texttt{deploy.mydsl} file (Step \circled{1}) and selecting Compile Deploy (Step \circled{2}) as shown in Figure~\ref{fig:cdeploy} generates a deployment packages.

\textit{Import user-interface project.} To import user-interface project, click on File Menu (Step \circled{1}), and select Import option (Step \circled{2}) as shown in Figure~\ref{fig:importlogic}.

\begin{figure*}[!ht]
\centering
\includegraphics[width=0.75\textwidth]{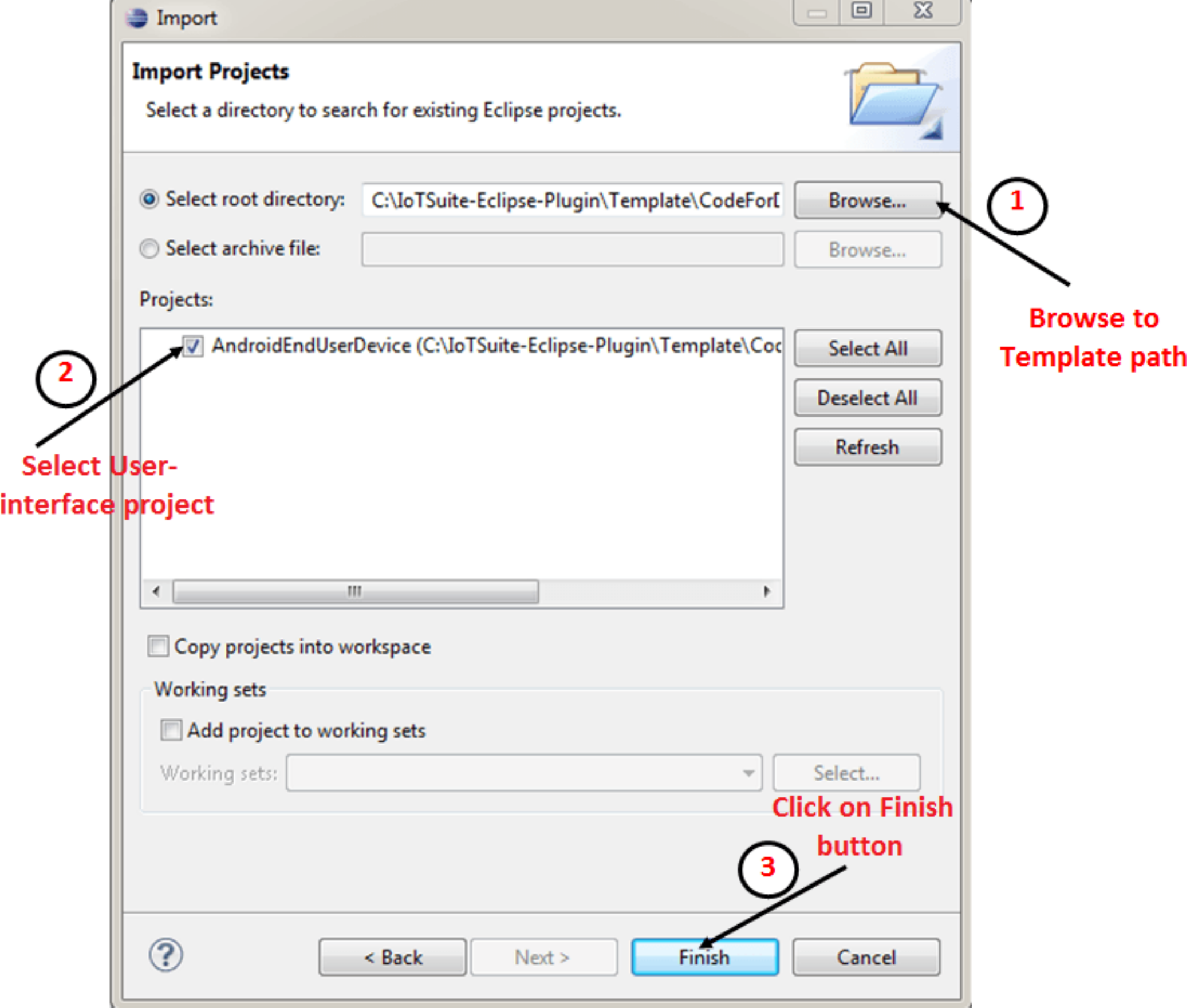}
\caption{Locate user-interface project}
\label{fig:eulocateui-1}
\end{figure*}

\textit{Locate user-interface project.} To locate user-interface project, browse to CodeForDeployment folder in Template path (Step \circled{1}), select project specified in the use-interaction specification (Step \circled{2}), and click on Finish button (Step \circled{3}) as shown in Figure \ref{fig:eulocateui-1}.

\begin{figure*}[!ht]
\centering
\includegraphics[width=0.82\textwidth]{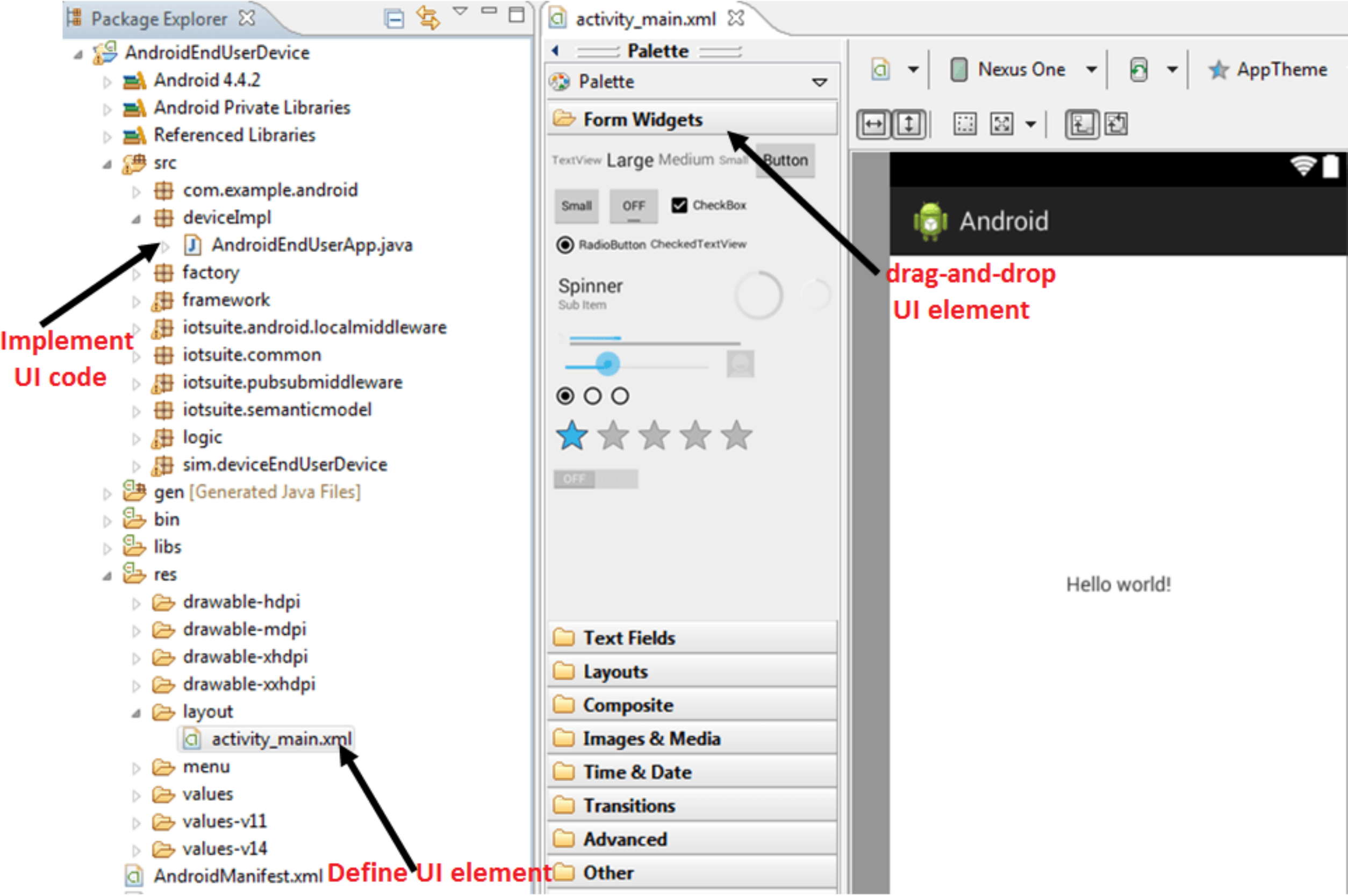}
\caption{Implementation of user-interface code}
\label{fig:implementui}
\end{figure*}

\textit{Implementing user-interface code.} In this step, developer implements user-interface code generated by compilation of user-interaction specification. Developer implements user-interface code in deviceImpl package~\cite[p.~15-16]{Patel201562}. The implementation of user-interface code involves the use of drag-and-drop functionality provided by form-widget using \texttt{activity\_main.xml} as shown in Figure \ref{fig:implementui}. The developer connects this interaction with generated framework
in the \texttt{AndroidEndUserApp.java} file. Listing \ref{enduser} shows how developer connects UI element with generated framework in deviceImpl package.

\lstset{language=Java, caption={The implementation of the generated \texttt{AndroidEndUserApp} class written by the developer.}, escapechar=\#, label=enduser}
\begin{lstlisting}
package deviceImpl;

import logic.*;
import framework.*;
import android.content.Context;
import android.app.Activity;
import iotsuite.android.localmiddleware.IDataListener;
import iotsuite.android.localmiddleware.PubSubsSensingFramework;

public class AndroidEndUserApp implements IEndUserApp, IDataListener {
	public static PubSubsSensingFramework pubSubSensingFramework;
	private Context appContext;
	public static Activity appActivity;
	public static String txtDisplay;
	
	public AndroidEndUserApp(Context context, LogicEndUserApp obj) {
		this.appContext = context;
		appActivity = (Activity) appContext;
		pubSubSensingFramework = PubSubsSensingFramework.getInstance();
		pubSubSensingFramework.registerForSensorData(this, "fireNotifyNotify");
		}

	@Override
	public void onDataReceived(String eventName, Object data) {
		// Developer connects UI element with generated framework 
		if (eventName.equals("fireNotifyNotify")) {
			FireStateStruct fireData = (FireStateStruct)data;
			txtDisplay= fireData.getfireValue();
		}
	}
}
\end{lstlisting}


\begin{figure*}[!ht]
\centering
\includegraphics[width=0.8\textwidth]{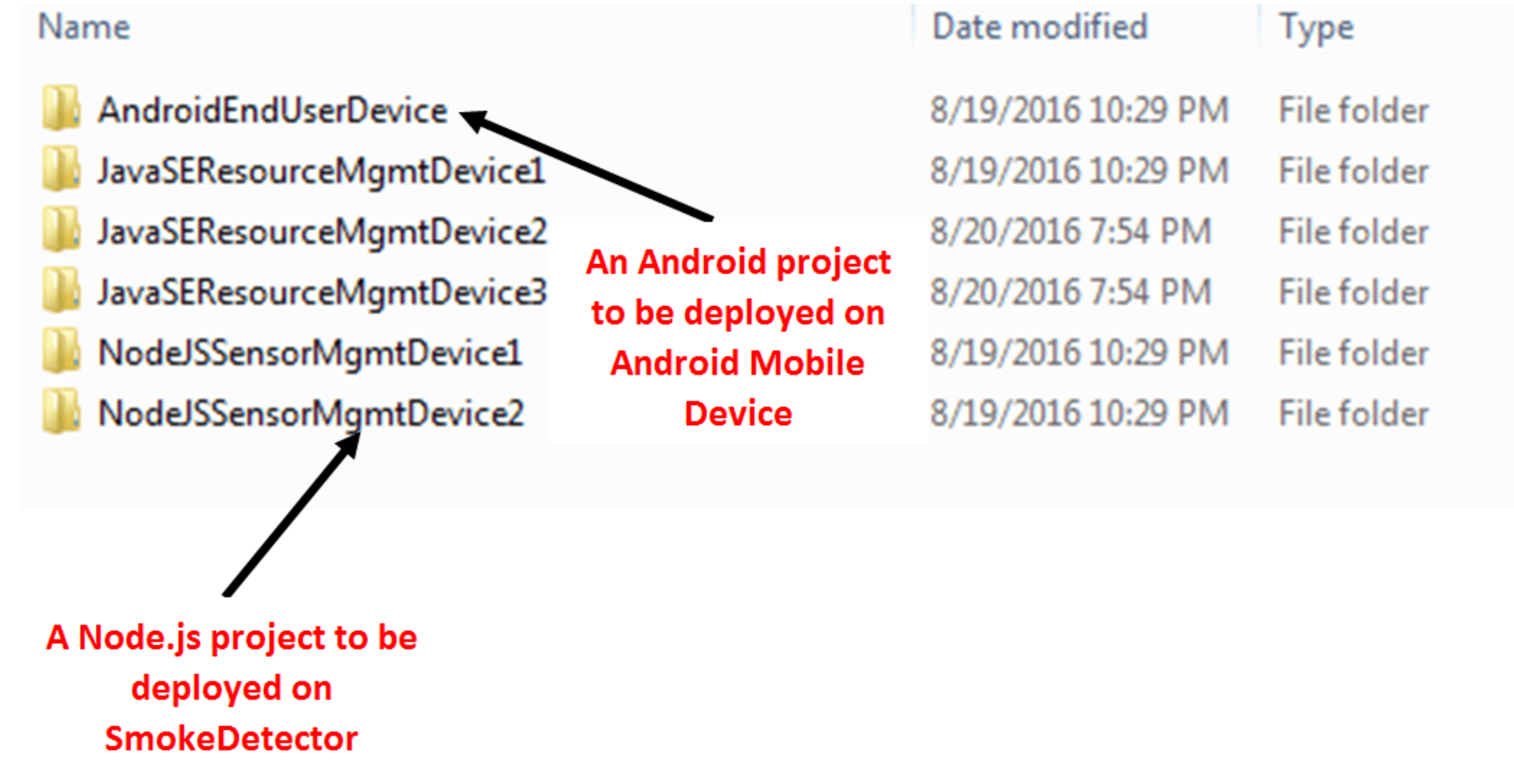}
\caption{Packages for target devices specified in the deployment spec.}
\label{fig:eudeploypackage}
\end{figure*}

\textit{Deployment of generated packages.} The output of compilation of deployment specification produce a set of platform
specific project/packages as shown in Figure~\ref{fig:eudeploypackage} for devices, specified in the
deployment specification (Refer Listing~\ref{eudeploydsl}). These projects compiled by device specific compiler designed for the target platform. The generated packages integrate the run-time system.

\subsection{Smart Home Monitoring Application}
\begin{figure}[!ht]
	\centering 
	\includegraphics[width=1.0\linewidth]{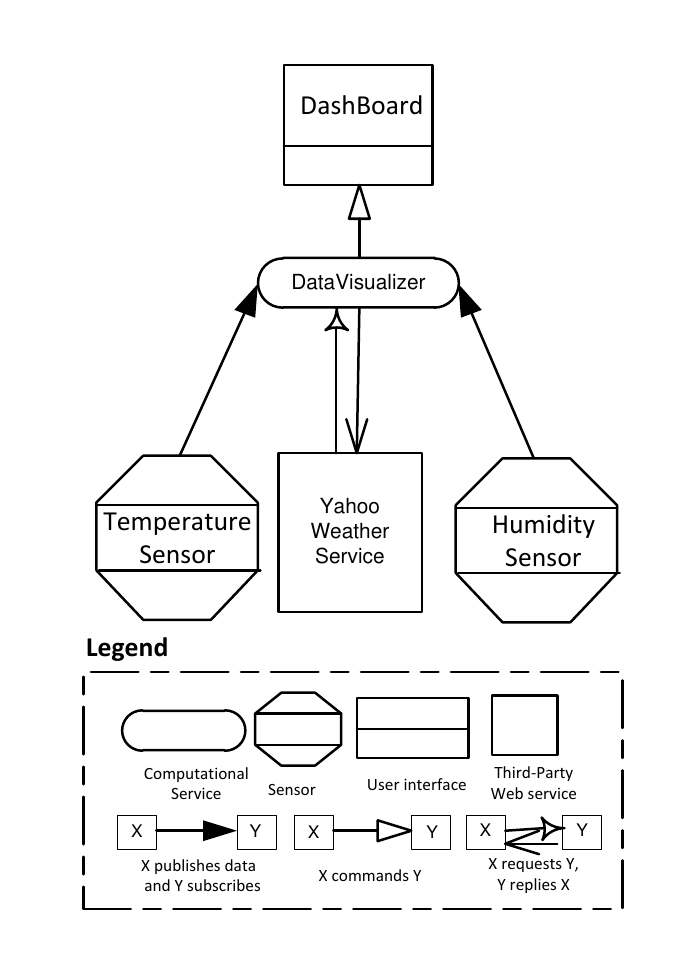} 
	\caption{Dataflow of Data Visualization Application} \label{fig:awarness}
\end{figure} 
A home consists of several rooms, each one is instrumented with several heterogeneous entities for providing resident's awareness.
To provide a resident's awareness, the system generates the current environment status on dashboard (e.g., humidity, temperature, outside temperature by interacting with external web services) as shown in Figure~\ref{fig:awarness}.

\textit{Specifying Domain specification.} Developer specifies domain specification using IoTSuite as shown in Listing~\ref{dvvocabDSL}. Each resource is charaterized by types of information it generates or consumes. A set of information is defined using the \texttt{structs} keyword (Listing~\ref{dvvocabDSL}, line~\ref{vocab:structs-start}). A set of sensors is declared using the \texttt{sensors} keyword (Listing \ref{dvvocabDSL}, line \ref{vocab:sensors-start}). Each sensor produces one or more sensor measurements along with the data-types specified in the data structure (Listing \ref{dvvocabDSL}, lines \ref{vocab:tsstructs-start}-\ref{vocab:tsstructs-end}), declared using the generate keyword. A periodic sensor samples results every d seconds for duration k seconds. For instance, a temperature sensor generates a \texttt{tempMeasurement} (Listing \ref{dvvocabDSL}, line \ref{vocab:generate}) of \texttt{TempStruct} type (Listing \ref{dvvocabDSL}, lines \ref{vocab:tsstructs-start}-\ref{vocab:tempstructs-end}). It samples data every 1 second for next 6 minutes (Listing \ref{dvvocabDSL}, line \ref{vocab:sample}). A request-based sensor responds its results only if it is requested. For instance, the \texttt{YahooWeatherService} provides temperature value of a location given by a \texttt{locationID} (Listing \ref{dvvocabDSL}, lines \ref{vocab:requestBasedSensors:start}-\ref{vocab:requestBasedSensors:end}).

\lstset{emph={String, generate, actuators, structs, double,  action, resources, long, storages, accessed, by, sample, period, for, eventDrivenSensors, onCondition, true, tags, =}, emphstyle={\color{blue}\bfseries\emph}, caption={Code snippet of domain spec.}, escapechar=\#, 
 label=dvvocabDSL}	
\begin{lstlisting}
structs:		#\label{vocab:structs-start}#																			
	TempStruct           #\label{vocab:tsstructs-start}#	                                                                   
		tempValue: double;                                                                  
		unitOfMeasurement : String;        #\label{vocab:tempstructs-end}#                                                 
	HumidityStruct                                                                          
		humidityValue : double;                                                             
		unitOfMeasurement :String;   #\label{vocab:tsstructs-end}#	                                                       
resources:                                                                                 
	sensors:        #\label{vocab:sensors-start}#	                                                                        
	   periodicSensors:                                                                    
	TemperatureSensor                                                               
		generate tempMeasurement:TempStruct;   #\label{vocab:generate}#	                                       
		sample period 1000 for 6000000;  #\label{vocab:sample}#	                                              
	HumiditySensor                                                                  
		generate humidityMeasurement:HumidityStruct;                                
		sample period 1000 for 6000000;                                              
 requestBasedSensors:    #\label{vocab:requestBasedSensors:start}#	                                                            
	YahooWeatherService                                                             
		generate weatherMeasurement:TempStruct accessed-by locationID : String; #\label{vocab:requestBasedSensors:end}#    
							
\end{lstlisting}


\textit{Compilation of domain specification.} To compile domain specification~(\texttt{vocab.mydsl} file), Right click on the \texttt{vocab.mydsl} file (Step \circled{1}) and click on Compile Vocab option (Step \circled{2}) as shown in Figure~\ref{fig:cvocab}.

\textit{Architecture specification.} Developer specifies architecture specification using IoTSuite as shown in Listing~\ref{dvarchdsl}. It is described as a set of computational services. It consists
of two types of computational services: (1) \texttt{Common} component specifies common operations (e.g., average, count,
sum ) in the application logic. (2) \texttt{Custom} specifies an application-specific logic. Additionally, each computational service is described by a set of inputs and outputs. For instance, the \texttt{DataVisualizer} consumes \texttt{tempMeasurement}, \texttt{humidityMeasurement}, and \texttt{weatherMeasurement} (Listing \ref{dvarchdsl}, lines~\ref{arch-consume-start}-\ref{arch-consume-end}), issues a \texttt{DisplaySensorMeasurement} command with a \texttt{sensorMeasurement} as an argument to \texttt{Dashboard} (line \ref{arch-command-end}).

\lstset{emph={generate, computationalServices, command, from, to, 
consume, request, Custom, Common, COMPUTE, AVG_BY_SAMPLE }, emphstyle={\color{blue}\bfseries\emph} }

\lstset{caption={A code snippet of architecture spec.}, escapechar=\#, label=dvarchdsl}
\begin{lstlisting}
computationalService:     
 Custom:                                                                 
	DataVisualizer                                                  
		consume tempMeasurement from TemperatureSensor;      #\label{arch-consume-start}#            
		consume humidityMeasurement from HumiditySensor;                
		consume weatherMeasurement from YahooWeatherService;          #\label{arch-consume-end}#   
		command DisplaySensorMeasurement(sensorMeasurement) to DashBoard; #\label{arch-command-end}#
\end{lstlisting}

\textit{Compilation of architecture specification.} To compile architecture specification~(\texttt{arch.mydsl} file), Right click on the \texttt{arch.mydsl} file (Step \circled{1}) and click on Compile Arch option (Step \circled{2}) as shown in Figure~\ref{fig:carch}.


\textit{Import application logic package.} To import application logic package, click on File Menu (Step \circled{1}), and select Import option (Step \circled{2}) as shown in Figure~\ref{fig:importlogic}.


\textit{Locate application logic package.} To locate application logic package, browse to Template path (Step \circled{1}), select application logic package (Step \circled{2}), and click on Finish button (Step \circled{3}) as shown in Figure~\ref{fig:locatelogic}.

\textit{Implement application logic.} The application logic proj-ect contains a generated framework that hides low-level details from a developer and allows the developer to focus on the application logic. The developer writes application
specific logic in logic package. Listing \ref{visualizer} shows generated Java programming framework from the architecture specification, defined in Listing~\ref{dvarchdsl}. 

To implement application logic of \texttt{DataVisualizer}, developers have to implement the generated abstract methods illustrated in Listing \ref{visualizer}. The \texttt{DataVisualizer} component consumes the \texttt{tempMeasurement}, \texttt{humidityMeasurement}, \texttt{weatherMeasurement} and issues \texttt{DisplaySensorMeasurement} command to \texttt{Dashboard} defined in Listing \ref{dvarchdsl}, lines~\ref{arch-consume-start}-\ref{arch-command-end}. To implement this logic, three methods (\texttt{onNewwe-atherMeasurement}, \texttt{onNewhumidityMeasurement}, and \texttt{onNewtempMeasurement}) have to implemented as shown in Listing~\ref{visualizer}. It reads \texttt{tempMeasurement} using \texttt{onNewtempMeasurement} method, \texttt{humidityMeasurement} using \texttt{onNewhumidityMeasurement} method, and \texttt{weatherMeasurement} using \texttt{onNewweatherMeasurement} method illustrated in Listing \ref{visualizer}. Additionally, it issues a \texttt{DisplaySensorMeasurement} command to \texttt{Dashboard} using \texttt{onNewweatherMeasurement} method as shown in Listing~\ref{visualizer}.
\lstset{language=Java, caption={The implementation of the Java
    abstract class \texttt{DataVisualizer} written by the developer.}, escapechar=\#, label=visualizer}
\begin{lstlisting}
package logic;

import iotsuite.pubsubmiddleware.PubSubMiddleware;
import android.content.Context;
import iotsuite.semanticmodel.*;
import framework.*;

public class LogicDataVisualizer extends DataVisualizer {

	double tempValue, humidityValue;

	public LogicDataVisualizer(PubSubMiddleware pubSubM, Device deviceInfo,
			Object ui, Context myContext) {
		super(pubSubM, deviceInfo);
	}

	@Override
	public void onNewweatherMeasurement(TempStruct arg) {
		// Read weatherMeasurement from YahooWeatherService
		// and set DisplaySensorMeasurement to Dashboard
		VisualizationStruct struct = new VisualizationStruct(tempValue,
				humidityValue, arg.gettempValue());
		DisplaySensorMeasurement(struct);
	}

	@Override
	public void onNewhumidityMeasurement(HumidityStruct arg) {
		// Read humidityMeasurement from HumiditySensor
		humidityValue = arg.gethumidityValue();
	}

	@Override
	public void onNewtempMeasurement(TempStruct arg) {
		// Read tempMeasurement from TemperatureSensor
		tempValue = arg.gettempValue();
	}
}
\end{lstlisting}

\textit{Userinteraction specification.} Developer specifies userinteraction specification using IoTSuite as shown in Listing~\ref{dvuidsl}. It defines what interactions are required by an application. We design a set of abstract interactors that denotes data exchange between an application and a user. \texttt{Notify} denotes information flow from an application to a user. For instance, an application notifies a user by providing temperature, humidity and outside temperature value by interacting with \texttt{YahooWeatherService}. It is declared using the \texttt{notify} keyword (Listing \ref{dvuidsl}, line \ref{ui-dashboard-start}). The application notifies users with the visualization information specified in the data structure (Listing \ref{dvuidsl}, lines \ref{ui-struct-start}-\ref{ui-struct-end}).

\lstset{emph={structs, resources,  userInteractions}, emphstyle={\color{blue}\bfseries\emph}%
}
\lstset{caption={Code snippet of user-interaction spec.}, label=dvuidsl}
\begin{lstlisting}
structs:																									
 VisualizationStruct        #\label{ui-struct-start}#                                                                             
  tempValue:double;                                                                                   
  humidityValue:double;                                                                               
  yahooTempValue:double;    #\label{ui-struct-end}#                                             
resources:                                                                                                 
 userInteractions:                                                                                       
  DashBoard                                                                                           
   notify DisplaySensorMeasurement(sensorMeasurement : VisualizationStruct) from DataVisualizer; #\label{ui-dashboard-start}#
\end{lstlisting}

\textit{Compilation of userinteraction specification.} To compile userinteraction specification~(\texttt{userinteraction.mydsl} file), Right click on the \texttt{userinteraction.mydsl} file (Step \circled{1}) and click on Compile UserInteraction option (Step \circled{2}) as shown in Figure~\ref{fig:eucui}.

\textit{Deployment specification.} Developer specifies deployment specification using IoTSuite as shown in Listing~\ref{dvdeploydsl}. It includes properties such as \texttt{location} that defines where a device is deployed, \texttt{resources} define component(s) to be deployed on a device, \texttt{language-platform} is used to generate an appropriate package for a device, \texttt{protocol} specifies a run-time system installed on a device to interact with other devices. Listing~\ref{deploydsl} shows a code snippet to illustrate these concepts. 
\texttt{SensorMgmtDevice} is located in room\#1 (line \ref{SensorMgmtDevice-room}),  \texttt{TemperatureSensor} and \texttt{HumditySensor} are attached with the device (line \ref{SensorMgmtDevice-resources}) and the device driver code for these two component is in \texttt{NodeJS} (line \ref{SensorMgmtDevice-platform}), \texttt{mqtt} runtime system is installed on \texttt{SensorMgmtDevice} (line \ref{SensorMgmtDevice-protocol}). 

\begin{figure*}[!ht]
\centering
\includegraphics[width=0.9\textwidth]{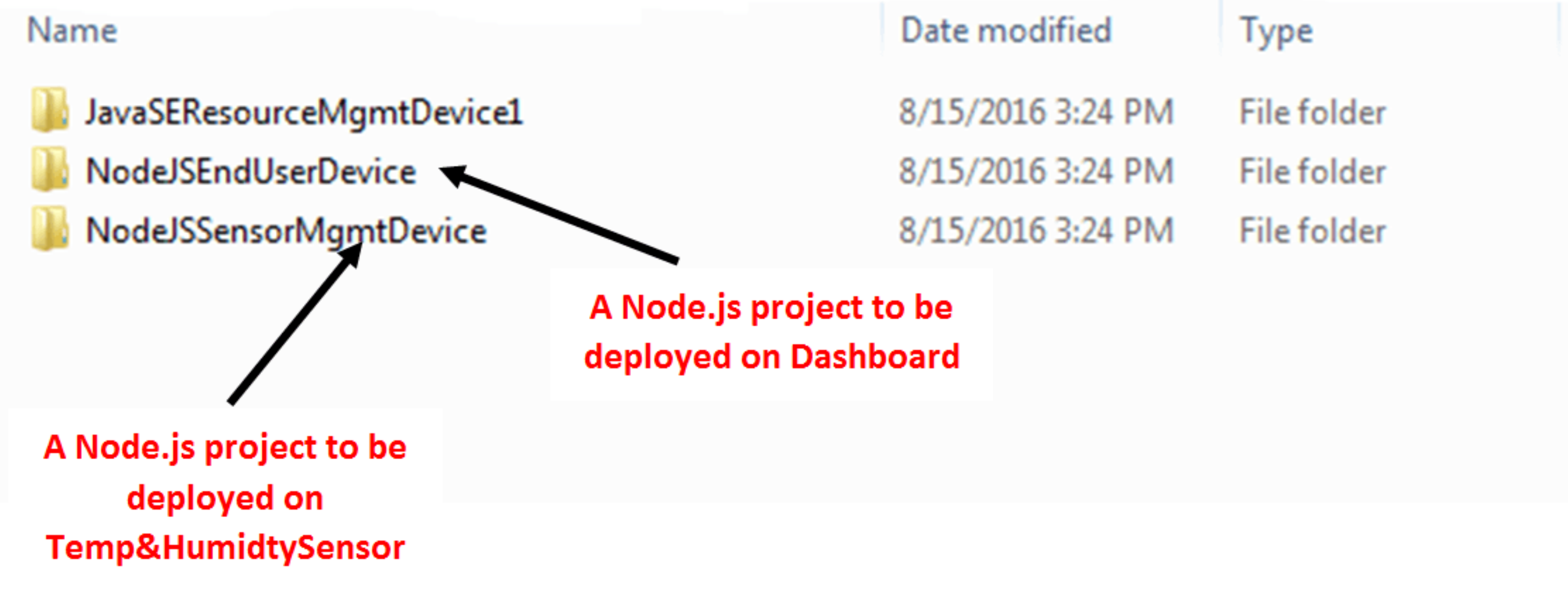}
\caption{Packages for target devices specified in the deployment spec.}
\label{fig:dvdeploypackage}
\end{figure*}

\lstset{emph={resources, softwarecomponents,  devices, location, protocol, description,database, language,platform}, emphstyle={\color{blue}\bfseries\emph}%
}
\lstset{caption={Code snippet of deployment spec.}, label=dvdeploydsl}
\begin{lstlisting}
devices:												
 SensorMgmtDevice:        #\label{SensorMgmtDevice-component}#                                             
	location:                                       
		Room:1 ;                   #\label{SensorMgmtDevice-room}#                   
	platform: NodeJS;                        #\label{SensorMgmtDevice-platform}#    
	resources: TemperatureSensor, HumiditySensor;  #\label{SensorMgmtDevice-resources}# 
	protocol: mqtt ;                  #\label{SensorMgmtDevice-protocol}#           
 ResourceMgmtDevice1:                                                
	location:                                       
	 	Room: 1 ;                                  
	platform: JavaSE;                           
	resources: ;                      
	protocol: mqtt  ;   
 EndUserDevice:                                                 
	location:                                       
		Room:1 ;                                   
	platform: NodeJS;                          
	resources: DashBoard;              
	protocol: mqtt ; 		                       
	 		                           
\end{lstlisting}

\textit{Compilation of deployment spec.} Right click on \texttt{deploy.mydsl} file (Step \circled{1}) and selecting Compile Deploy (Step \circled{2}) as shown in Figure~\ref{fig:cdeploy} generates a deployment packages.

\textit{Import user-interface project.} To import user-interface project, click on File Menu (Step \circled{1}), and select Import option (Step \circled{2}) as shown in Figure~\ref{fig:importlogic}.

\textit{Locate user-interface project} To locate user-interface project, browse to CodeForDeployment folder in Template path (Step \circled{1}), select project specified in the use-interaction specification (Step \circled{2}), and click on Finish button (Step \circled{3}) as shown in Figure \ref{fig:eulocateui-1}.

\textit{Implementing user-interface code.} In this step, developer implements user-interface code generated by compilation of user-interaction specification. Developer implements user-interface code in deviceImpl package~\cite[p.~15-16]{Patel201562}. To reduce development efforts to implement functionality of Dashboard, we are generating code with default template.

\textit{Deployment of generated packages.} The output of compilation of deployment specification produce a set of platform
specific project/packages as shown in Figure~\ref{fig:dvdeploypackage} for devices, specified in the
deployment specification (Refer Listing~\ref{dvdeploydsl}). These projects compiled by device specific compiler designed for the target platform. The generated packages integrate the run-time system.
\section{Conclusion and Future work}\label{sec:conclusion}

Since the main goal of this research is to make IoT application development easy 
for stakeholders, we believe that our IoT application development process should be supported by 
tools to be applicable in an effective way. Therefore, this paper introduces a design and implementation 
of IoTSuite, a suite of tools, for reducing burden of each phase of IoT application development process.
Moreover, we take different class of IoT applications, largely found in the IoT literature, and demonstrate these IoT application development using IoTSuite. 
These applications have been tested on several IoT technologies such as Android, Raspberry PI, Arduino, and JavaSE-enabled devices, Messaging protocols such as MQTT, CoAP, WebSocket, Server technologies such as Node.js, Relational database such as MySQL, and Microsoft Azure Cloud services.

It consists of the following components to aid stakeholders: (1) We integrate a customized editor 
support for specifying high-level textual specification with the facilities of syntax
coloring and syntax error reporting. (2) A compiler parses the high-level specifications and supports the
application development phase by producing a programming framework that reduces development effort.
This compiler knows at run-time which code generation plug-ins are installed and generates code
in a  target implementation language. (3) A deployment module is supported by mapping and linking 
techniques. They together support the deployment phase by producing device-specific code to result in a
distributed system collaboratively hosted by individual devices. (4) A runtime system leverages existing 
middleware platforms and generates glue code to customize them with respect to needs of IoT applications. This 
mechanism increases the possibility of executing applications on different middleware.

\section{Future work}\label{sec:futurework}

This work presents steps involved in IoT application development, 
and prepares a  foundation  for  our future research work. 
Our future work will proceed in the following complementary directions, discussed below.

\fakeparagraph{\emph{Testing support for IoT application development}}
Our near term future work will be  to provide support for the testing phase. A key advantage of 
testing is that it emulates the execution of an application before deployment so as to identify 
possible conflicts, thus reducing application debugging effort. The support will be provided 
by integrating an open source simulator in IoTSuite. This simulator will enable transparent 
testing of IoT applications in a simulated physical environment. Moreover, we expect to enable the simulation of 
a hybrid environment, combining both real and physical entities. Currently, we are investigating 
open source simulators for IoT applications. We see Siafu\footnote{\url{http://siafusimulator.org/}} 
as a possible candidate due to its open source and thorough documentation.

\fakeparagraph{\emph{Mapping algorithms cognizant of heterogeneity}}
We will provide rich abstractions to express both the
properties of the devices (e.g., processing and storage
capacity, networks it is attached to, as well as monetary cost
of hosting a computational service), as well as the requirements
from stakeholders regarding the preferred placement of the
computational services of the applications. These will then be
used to guide the design of algorithms for efficient mapping
(and possibly migration) of computational services on devices.

\bibliographystyle{elsarticle-harv}
\bibliography{main}

\end{document}